# Higher Order Elastic Instabilities of Metals: From Atom to Continuum Level

Kun Wang[a], Jun Chen[a,b*], Wenjun Zhu[c], Meizhen Xiang[a]

[a] *Laboratory of Computational Physics, Institute of Applied Physics and Computational Mathematics, Beijing 100088, PR China*
[b] *Center for Applied Physics and Technology, Peking University, Beijing 100071, China*
[c] *National Key Laboratory of Shock Wave and Detonation Physics, Institute of Fluid Physics, Mianyang 621900, China*

## ABSTRACT

Strain-based theory on elastic instabilities is being widely employed for studying onset of plasticity, phase transition or melting in crystals. And size effects, observed in nano-materials or solids under dynamic loadings, needs to account for contributions from strain gradient. However, the strain gradient based higher order elastic theories on the elastic instabilities are not well established to enable one to predict high order instabilities of solids directly at atom level. In present work, a general continuum theory for higher order elastic instabilities is established and justified by developing an equivalent description at atom level. Our results show that mechanical instability of solids, triggered by either strain or strain gradient, is determined by a simple stability condition consisting of strain or strain gradient related elastic constants. With the atom-level description of the higher order elasticity, the strain-gradient elastic constants could be directly obtained by a molecular statics procedure and then serve as inputs of the stability condition. In this way, mechanical instabilities of three metals, i.e., copper, aluminum and iron, are predicted. Alternatively, ramp compression technique by nonequilibrium molecular dynamics (NEMD) simulations is employed to study the higher order instabilities of the three metals. The predicted critical strains at onset of instabilities agree well with the results from the NEMD simulations for all the metals. Since the only inputs for the established higher order elastic theory are the same as atomic simulations, i.e., atomic potentials and structures of solids, the established theory is completely equivalent to empirical-potential based atomic simulations methods, at least, for crystals.

## 1. Introduction

Elastic instability has found widespread applications in predicting onset of plasticity, phase transition and melting for varieties of materials. It is known that size and microstructure effects cannot be averaged out in macroscopic response when the relative macroscopic dimension is comparable to the characteristic size of the microstructures. Such situation is often encountered in nano-materials. The same could also occur in shocked materials where width of wave front due to the shock is comparable to the microstructure (lattice constants or microscopic characteristic

---

length). These effects could be successfully captured in continuum theories that involve a material length scale, such as strain gradient related theories. The first strain gradient elasticity theory is proposed by Mindlin (Mindlin and Eshel, 1968; Mindlin, 1965), and later developed by other authors (Fleck and Hutchinson, 1997; Fleck et al., 1994; Hadjesfandiari and Dargush, 2011; Lam et al., 2003; Polyzos and Fotiadis, 2012; Yang et al., 2002). Specially, couple stress theory proposed by (Toupin, 1964) is another kind of higher order continuum theory, which uses higher order rotation gradients or curvature tensor as deformation metrics. While the original strain gradient elasticity is formulated in terms of the first and the second derivatives of displacement. The couple stress theory is found to be a special case of the strain gradient theory through neglecting effects of dilatation and deviatoric stretch (Fleck and Hutchinson, 1997; Lam et al., 2003; Yang et al., 2002). Because strain gradient has eighteen independent components, a large number of material parameters (strain-gradient elastic constants), work-conjugates to strain gradient, emerge in the strain gradient theory. To reduce the number of these unknown parameters, elastic solids are often assumed to be linear isotropic, and thus only five additional independent material parameters need to be considered when compared with conventional strain-based elastic theory (Fleck and Hutchinson, 1997). Due to the symmetric character of couple stress tensor constraint by higher-order equilibrium condition (Yang et al., 2002), (Lam et al., 2003) find that the skew-symmetric part of rotation gradient does not contribute to the deformation energy which leads to a reduction of the number of independent material parameters from five to three. Alternatively, (Hadjesfandiari and Dargush, 2011) establishes a skew-symmetric character of the couple stress tensor by considering true continuum kinematical displacement and rotation. Although extensive literatures on strain gradient elasticity have now appeared, the theory still appears to be phenomenological because the additional material parameters are unknown and are often determined by fitting to some well-known analytic results or experiments. These determination methods seem to be the most effective for strain gradient plasticity, but not for elasticity since the latter has exact microscopic lattice model (Mindlin, 1972; Polyzos and Fotiadis, 2012). (DiVincenzo, 1986; Maranganti and Sharma, 2007; Stengel, 2013, 2016) establishes a lattice dynamic approach to determine the strain-gradient elastic constants. Key point of the approach is to numerically fitting phonon dispersion relations along certain high symmetry directions in order to acquire the requisite elastic constants. However, great cautions should be taken when performing the fitting procedures. For example, the fit should be carried out starting from k-vectors in the vicinity of zero to the one where dispersive effects just start to kick in (Maranganti and Sharma, 2007). Fitting at k-vector, corresponding to regions where frequencies are very high and dispersive effects are large, will results in spurious estimations of the elastic constants. Alternatively, a statistical mechanics approach, which relates the strain-gradient elastic constants to atomic displacement correlations in a molecular dynamics (MD) ensemble, is proposed by (Maranganti and Sharma, 2007). This approach could be also applied for estimating the elastic constants of non-crystalline systems. To acquire the elastic constants with high precision, a large simulation cell should be employed in the MD simulations. Both of the two approaches adopt a "dynamic" or "indirect" (statistical) way to acquire a proper estimation of the elastic constants. In contrast, we will present a static (direct) way to calculate the strain-gradient elastic constants in this work. The static approach also enables us to calculate the higher order stresses which have no yet been determined directly at atom level so far. Additionally, strain gradient is found to be responsible to mechanical instabilities of metals under extreme strain rates (Wang et al., 2017). However, traditional theories on the mechanical instabilities are strain-based theories. For such reason, we formulate an elastic stability condition which encompasses effects of both strain and strain gradient. The stability condition could enable one to study not only mechanical instabilities of real solids, but also strength of coupling effects between strain and strain gradient. Because in some multiple-scale numerical methods, for example phase field method (Levitas and Preston, 2002a, b; Tröster et al., 2014; Tröster et al., 2002), equilibrium equation and stability condition are the bases for constructing constitutive relationships of materials. Specially, phase transitions between two phases under high pressure often involve deformation with large strains. Thereby, equilibrium equation and stability condition are derived for both small and finite strains in this work.

The remainder of present work is organized as follows. In Part 2, we first demonstrate independent deformation metrics adopted in this work is equivalent to the original ones defined by Mindlin. Then equilibrium equation and stability condition are derived for small strains in Part 3, and for finite strains in Part 4. Microscopic approaches to the elastic constants defined in our theory are given in Part 5. With the stability condition, we evaluate critical strain, at which crystals begin to become instable, for copper, aluminum and iron through the microscopic approaches, and make a direct comparison with results from nonequilibrium molecular dynamic (NEMD) simulations in Part 6 in order to check the present theory. It should be noted that no fitting procedures or uncertain parameters are involved during this process. And, finally, we end this work by concluding in Part 7. By convention, bold font letter stands for vector or tensor and otherwise, it denotes magnitude of the corresponding vector or merely a scalar. If not specified, summation over repeated indexes is employed.

## 2. Independent Deformation Metrics

Equilibrium equation and boundary conditions could be established through the principle of virtual work which should be written in terms of independent deformation metrics. Thereby, before establishing equilibrium equation, as well as stability condition, for solids, independent strain metrics must be identified. Following traditional definitions of small strain and rotation tensors, the first derivative of displacement ($\boldsymbol{u}$) is written as a sum of the strain ($\boldsymbol{\varepsilon}$) and rotation ($\boldsymbol{\omega}$), that is,

$$u_{i,j} = \varepsilon_{ij} + \omega_{ij}, \tag{1}$$

where

$$\varepsilon_{ij} = \frac{1}{2}(u_{i,j} + u_{j,i}); \ \omega_{ij} = \frac{1}{2}(u_{i,j} - u_{j,i}). \tag{2}$$

Then the second derivative of the displacement is

$$\kappa_{ijk} \stackrel{\text{def}}{=} u_{k,ij} = \varepsilon_{ki,j} + \omega_{ki,j}, \tag{3}$$

where we have used notation $(\cdots)_{,i}$ to represent $\partial(\cdots)/\partial x_i$, $x_i$ is the $i$-th Cartesian component of position vector. Variation of total strain energy in volume $V$ due to a small virtual variation of displacement ($\delta\boldsymbol{u}$) is

$$\delta \overline{W} = \int_V \delta w \, dV = \int_V \left(\sigma_{ij}\delta\varepsilon_{ij} + \tilde{\tau}_{ijk}\delta\kappa_{ijk}\right)dV, \tag{4}$$

where $\delta\boldsymbol{\varepsilon}$ and $\delta\boldsymbol{\kappa}$ are determined by the variation of displacement field $\delta\boldsymbol{u}$ (as well as boundary condition). Stress ($\boldsymbol{\sigma}$) and higher order stress ($\tilde{\boldsymbol{\tau}}$) are defined as work conjugated to $\boldsymbol{\varepsilon}$ and $\boldsymbol{\kappa}$. Mindlin (1965) has established his equilibrium equation and boundary conditions from Eq. (4), while Lam and et. al. (Lam et al., 2003) propose new equilibrium equation and boundary conditions in terms of new independent stratified metrics. In present work, we will propose another independent metrics which are equivalent to that of Lam, and more convenient for studying elastic stabilities of solids.

According to Eq. (3), $\kappa_{ijk}$ has eighteen independent components in considering that the second order mixed derivatives do not rely on the sequence of the related variables. However, in the right hand side of Eq. (4), the independent components are eighteen for $\varepsilon_{ki,j}$ and nine for $\omega_{ki,j}$ after considering permutation symmetry and skew-symmetry between index $i$ and $k$ held by $\varepsilon_{ki,j}$ and $\omega_{ki,j}$, respectively. That is to say, Eq. (4) is over-/under-determined. This is because that the components of strain gradient ($\varepsilon_{ki,j}$) and rotation gradient ($\omega_{ki,j}$) relate to each other. The relationship between rotation gradient and strain gradient is found to be (Verification is given in Appendix A)

$$\omega_{ij,k} = \varepsilon_{ki,j} - \varepsilon_{jk,i}. \tag{5}$$

Thus, by utilizing Eq. (3), the virtual variation of the strain energy density ($w$) could be expressed as

$$\begin{aligned}\delta w &= \sigma_{ij}\delta\varepsilon_{ij} + \tilde{\tau}_{ijk}\delta\kappa_{ijk} = \sigma_{ij}\delta\varepsilon_{ij} + \tilde{\tau}_{ijk}\big(\delta\varepsilon_{ki,j} + \delta\omega_{ki,j}\big) \\ &= \sigma_{ij}\delta\varepsilon_{ij} + \tilde{\tau}_{ijk}\big(\delta\varepsilon_{ki,j} + \delta\varepsilon_{jk,i} - \delta\varepsilon_{ij,k}\big) = \sigma_{ij}\delta\varepsilon_{ij} + \tau_{ijk}\delta\varepsilon_{ij,k},\end{aligned} \tag{6}$$

where

$$\tau_{ijk} = \tilde{\tau}_{kij} + \tilde{\tau}_{jki} - \tilde{\tau}_{ijk}. \tag{7}$$

If we take the independent components of strain gradient as deformation metrics, then $\boldsymbol{\tau}$ is work conjugated with strain gradient, which relates to the traditional higher order stress by (7). Below, we will manifest that traditional couple stress theory is a special case of the strain gradient theory under the new deformation metrics.

Rotation vector is defined by

$$\theta_i = \frac{1}{2}\epsilon_{ijk}\omega_{jk}, \tag{8}$$

where $\epsilon_{ijk}$ is alternating tensor. The rotation vector relates to the rotation tensor by

$$\omega_{ij} = \epsilon_{ijk}\theta_k. \tag{9}$$

In the couple stress theory, virtual variation of strain energy density is

$$\delta w = \sigma_{ij}\delta\varepsilon_{ij} + m_{ji}\delta\chi_{ij}, \tag{10}$$

where curvature tensor ($\chi_{ij}$) is

$$\chi_{ij} = \theta_{i,j} = \frac{1}{2}\epsilon_{ikl}\omega_{kl,j} \tag{11}$$

and $m_{ji}$ is work conjugated with $\chi_{ij}$. Substituting Eq. (5) into (11) and further substituting the obtained result into Eq. (10), we get

$$\begin{aligned}\delta w &= \sigma_{ij}\delta\varepsilon_{ij} + \frac{1}{2}m_{jl}\epsilon_{lki}\big(\delta\varepsilon_{jk,i} - \delta\varepsilon_{ij,k}\big) \\ &= \sigma_{ij}\delta\varepsilon_{ij} + \frac{1}{2}\big(m_{il}\epsilon_{ljk} - m_{jl}\epsilon_{lki}\big)\delta\varepsilon_{ij,k}.\end{aligned} \tag{12}$$

This is consistent with Eq. (6) when

$$\tau_{ijk} = \frac{1}{2}\big(m_{il}\epsilon_{ljk} - m_{jl}\epsilon_{lki}\big). \tag{13}$$

It is worth noting that all terms containing $\delta\varepsilon_{ij,j}$ (repeated indexes are not summed here) vanish in Eq. (12). From Eq. (5) and (9), we have
$$\delta\varepsilon_{ij,j} = \delta\varepsilon_{jj,i} + \delta\omega_{ij,j}. \tag{14}$$
The first term in the right hand side of Eq. (14) represents dilatation gradient and the second term is deviatoric part of the curvature tensor. Thereby, the strain gradient theory reduces to the couple stress theory when contributions of the dilatation gradient and the deviatoric stretch gradient to strain energy density are neglected. Similar conclusion is also arrived by Lam and et. al (Lam et al., 2003) through decomposing the symmetric second-order deformation gradient into trace and traceless part. Since strain gradient theory represented by strain and strain gradient is compatible the well justified couple stress theory (Hadjesfandiari and Dargush, 2011; Yang et al., 2002), we will using this representation described in this part to establish equilibrium equation and stability conditions for solids in the next two parts.

## 3. Strain-Gradient Related Stability Criteria for Small Strains

For a solid at fixed temperature and volume, mechanical equilibrium state is reached through "strain control" (Morris Jr and Krenn, 2000). Mathematically, the equilibrium state requires that Helmholtz free energy ($\bar{F}$) of the solid within the volume ($V$) reach, at least, its local minimum value with respect to reconfiguration under fixed boundary condition $\int \delta u_i \hat{n}_i dS = 0$, where $\delta u_i$ represents infinitesimal virtual variation of displacement and $\hat{n}_i$ is outer normal to surface ($S$) of the volume. This is equivalent to require that
$$\delta\bar{G} = \delta\bar{F} - \delta\bar{W} = 0, \tag{15}$$
where $\bar{W}$ is the work done by external force ($t_{ij}$) through the surface. That is,
$$\delta\bar{W} = \int t_{ij}\hat{n}_j \delta u_i dS = \int t_{ij}\delta u_{i,j} dV = \int (t_{ij}\delta\varepsilon_{ij} + t_{ij}\delta\omega_{ij})dV. \tag{16}$$
where $\delta\varepsilon_{ij} = \frac{1}{2}(\delta u_{i,j} + \delta u_{j,i})$ is incremental strain, and $\delta\omega_{ij} = \frac{1}{2}(\delta u_{i,j} - \delta u_{j,i})$ is incremental rotation. According to Eq. (16), $t_{ij}$ is positive when direction of the external force is along $\hat{\mathbf{n}}$. In above derivations, divergence theorem has been used. According to (Mindlin, 1965) and (DiVincenzo, 1986), potential energy density of elastic medium could be expanded in terms of the first and higher order derivatives of displacement gradient. In part 1, we have demonstrated that this representation is equivalent to the one where the derivatives of displacement gradient are replaced by strain and its gradients. It should be noted that the potential energy density only relies on "states" represented by strain and its gradients, instead of displacement and its derivatives, after the replacement. Thus the strain gradient could be conveniently handled in thermodynamics like other state variables, for example strain or specific volume. Let $f$ to be the Helmholtz free energy per unit volume, then the total free energy in the volume ($V$) surrounded by a piecewise smooth surface ($S$) is
$$\delta\bar{F} = \int \delta f(\boldsymbol{\varepsilon}, \nabla\boldsymbol{\varepsilon}, \boldsymbol{X}) dV = \int \left(\frac{\partial f}{\partial \varepsilon_{ij}}\delta\varepsilon_{ij} + \frac{\partial f}{\partial \varepsilon_{ij,k}}\delta\varepsilon_{ij,k}\right) dV$$
$$= \int \left[\frac{\partial f}{\partial \varepsilon_{ij}}\delta\varepsilon_{ij} - \left(\frac{\partial f}{\partial \varepsilon_{ij,k}}\right)_{,k}\delta\varepsilon_{ij}\right] dV + \int \frac{\partial f}{\partial \varepsilon_{ij,k}}\delta\varepsilon_{ij}\hat{n}_k dS. \tag{17}$$
where and $\boldsymbol{\varepsilon}$ and $\nabla\boldsymbol{\varepsilon}$ denote strain and strain gradient, respectively. Substituting (16) and (17) into (15), we get
$$\int \left[\frac{\partial f}{\partial \varepsilon_{ij}} - \left(\frac{\partial f}{\partial \varepsilon_{ij,k}}\right)_{,k} - t_{ij}\right]\delta\varepsilon_{ij} dV - \int t_{ij}\delta\omega_{ij} dV + \int \frac{\partial f}{\partial \varepsilon_{ij,k}}\delta\varepsilon_{ij}\hat{n}_k dS = 0. \tag{18}$$

Supposing that $\frac{\partial f}{\partial \varepsilon_{ij,k}}$ are constants along the surface for an equilibrium state (Mindlin and Eshel, 1968), it is then straight wards to shown that the last integral of equation (18) is zero by using the fixed boundary condition. Because at equilibrium states, equation (18) must be satisfied for any given $\delta\varepsilon_{ij}$ and $\delta\omega_{ij}$. Thereby, we have

$$\frac{\partial f}{\partial \varepsilon_{ij}} - \left(\frac{\partial f}{\partial \varepsilon_{ij,k}}\right)_{,k} - t_{ij} = 0, \tag{19a}$$

or

$$t_{ij} = \frac{\partial f}{\partial \varepsilon_{ij}} - \left(\frac{\partial f}{\partial \varepsilon_{ij,k}}\right)_{,k} = \frac{1}{V(X)}\frac{\partial \bar{F}}{\partial \varepsilon_{ij}} - \left(\frac{1}{V(X)}\frac{\partial \bar{F}}{\partial \varepsilon_{ij,k}}\right)_{,k}, \tag{19b}$$

and

$$t_{ij}\delta\omega_{ij} = 0. \tag{20}$$

Equation (19b) could be taken as a definition of Cauchy stresses at configuration $\{X\}$. Since the small strain $\varepsilon_{ij}$, as well as $\varepsilon_{ij,k_k}$, is symmetric with respect to $i$ and $j$, $t_{ij}$ should be a symmetric tensor which naturally satisfies the equation (20).

An equilibrium state, satisfying condition (15), is stable when

$$\delta^2 \bar{G} = \int \delta^2 f(\boldsymbol{\varepsilon}, \nabla\boldsymbol{\varepsilon}, T)dV > 0, \tag{21}$$

where

$$\delta^2 f(\boldsymbol{\varepsilon}, \nabla\boldsymbol{\varepsilon}, T) = \frac{\partial^2 f}{\partial \varepsilon_{ij}\partial \varepsilon_{kl}}\delta\varepsilon_{ij}\delta\varepsilon_{kl} + \frac{\partial^2 f}{\partial \varepsilon_{ij}\partial \varepsilon_{kl,n}}\delta\varepsilon_{ij}\delta\varepsilon_{kl,n} + \frac{\partial^2 f}{\partial \varepsilon_{ij,m}\partial \varepsilon_{kl}}\delta\varepsilon_{ij,m}\delta\varepsilon_{kl}$$
$$+ \frac{\partial^2 f}{\partial \varepsilon_{ij,m}\partial \varepsilon_{kl,n}}\delta\varepsilon_{ij,m}\delta\varepsilon_{kl,n} \tag{22}$$

Substituting (22) into (21), and after integrating by part, we will get

$$\delta^2 \bar{G} = \int \left(\frac{\partial^2 f}{\partial \varepsilon_{ij}\partial \varepsilon_{kl}} - \left(\frac{\partial^2 f}{\partial \varepsilon_{ij}\partial \varepsilon_{kl,n}}\right)_{,n}\right)\delta\varepsilon_{ij}\delta\varepsilon_{kl}dV + \int \frac{\partial^2 f}{\partial \varepsilon_{ij,m}\partial \varepsilon_{kl,n}}\delta\varepsilon_{ij,m}\delta\varepsilon_{kl,n}dV$$
$$+ \int \frac{\partial^2 f}{\partial \varepsilon_{ij}\partial \varepsilon_{kl,n}}\delta\varepsilon_{ij}\delta\varepsilon_{kl}\hat{n}_n dS > 0 \tag{23}$$

From finite-strain continuum elasticity theory (Thurston and R., 1964; Wallace, 1970), the Helmholtz free energy could be expanded into serials of finite strains. Here, we extend the ideas to include influences of strain gradients on the Helmholtz free energy. To understand the roles of strain gradients, one may image a stationary state sustained via certain nonequilibrium processes, for example materials at elastic wave front under linear ramp compressions. Thus, $\bar{F}$ could be expanded into serials, to the second order, in terms of temperature ($T$), $\boldsymbol{\varepsilon}$ and $\nabla\boldsymbol{\varepsilon}$, with respect to certain reference states, namely,

$$\rho\bar{F}(\boldsymbol{\varepsilon}, \nabla\boldsymbol{\varepsilon}, T) = -\rho\bar{F}(\mathbf{0}, \mathbf{0}, T) + \sigma_{ij}\varepsilon_{ij} + \tau_{ijm}\varepsilon_{ij,m} + \frac{1}{2}C_{ijkl}\varepsilon_{ij}\varepsilon_{kl} + W_{ijkln}\varepsilon_{ij}\varepsilon_{kl,n}$$
$$+ \frac{1}{2}\Pi_{ijmkln}\varepsilon_{ij,m}\varepsilon_{kl,n}, \tag{24}$$

where $\rho = 1/V(X)$. The coefficients in above equation are defined as energy-conjugates to strains, strain gradients and their second order mixture, that is,

$$\sigma_{ij} = \frac{1}{V(X)}\frac{\partial \bar{F}}{\partial \varepsilon_{ij}}, \tag{25}$$

$$C_{ijkl} = \frac{1}{V(X)} \frac{\partial^2 \bar{F}}{\partial \varepsilon_{ij} \partial \varepsilon_{kl}}, \tag{26}$$

$$\tau_{ijm} = \frac{1}{V(X)} \frac{\partial \bar{F}}{\partial \varepsilon_{ij,m}}, \tag{27}$$

$$W_{ijkln} = \frac{1}{V(X)} \frac{\partial^2 \bar{F}}{\partial \varepsilon_{ij} \partial \varepsilon_{kl,n}}, \tag{28}$$

$$\Pi_{ijmkln} = \frac{1}{V(X)} \frac{\partial^2 \bar{F}}{\partial \varepsilon_{ij,m} \partial \varepsilon_{kl,n}}, \tag{39}$$

Generally speaking, the other elastic constants, energy-conjugates to strain and strain gradient, are a function of strain and temperature at the reference states when the strain gradients are not very large. A detailed example will be given later this work. Then condition (23) becomes

$$\delta^2 \bar{G} = \int (C_{ijkl} - W_{ijkln,n}) \delta \varepsilon_{ij} \delta \varepsilon_{kl} dV + \int \Pi_{ijmkln} \delta \varepsilon_{ij,m} \delta \varepsilon_{kl,n} dV > 0, \tag{30}$$

where the surface contribution are neglected since only bulk properties of solids are considered here. The second integral of the last inquality in (30) could by further expressed in terms of $\delta \varepsilon_{ij} \delta \varepsilon_{kl}$ through procedures below:

$$\int \Pi_{ijmkln} \delta \varepsilon_{ij,m} \delta \varepsilon_{kl,n} dV = -\int (\Pi_{ijmkln} \delta \varepsilon_{ij,m})_{,n} \delta \varepsilon_{kl} dV + \int \Pi_{ijmkln} \delta \varepsilon_{ij,m} \delta \varepsilon_{kl} \hat{n}_n dS$$

$$= -\int \Pi_{ijmkln,n} \delta \varepsilon_{ij,m} \delta \varepsilon_{kl} dV - \int \Pi_{ijmkln} \delta \varepsilon_{ij,mn} \delta \varepsilon_{kl} dV + \int \Pi_{ijmkln} \delta \varepsilon_{ij,m} \delta \varepsilon_{kl} \hat{n}_n dS \tag{31}$$

It is easy to demonstrate that the first term of the last equality in above equation represents strain energy contributed by the higher order stresses by using Eq. (70) and (68d) given in the next part. And this term is the major part that determines strain gradient stability. That is to say, condition of the strain gradient stability requires $\Pi_{ijmkln,n}$ to be negative definite. This is consistent with our recent interpretations on the instability of iron before phase transition under extreme strain rates (Wang et al., 2017). In this work, we intend to generalize the result for plastic metals and, most importantly, give a reasonable estimation of relative contributions of strain and strain gradient to the stabilities of solids. The first term of the last equality in Eq. (31) could be calculated by part integral further, that is,

$$\int \Pi_{ijmkln,n} \delta \varepsilon_{ij,m} \delta \varepsilon_{kl} dV = -\int (\Pi_{ijmkln,n} \delta \varepsilon_{kl})_{,m} \delta \varepsilon_{ij} dV + \int \Pi_{ijmkln,n} \delta \varepsilon_{ij} \delta \varepsilon_{kl} \hat{n}_m dS$$

$$= -\int \Pi_{ijmkln,mn} \delta \varepsilon_{kl} \delta \varepsilon_{ij} dV - \int \Pi_{ijmkln,n} \delta \varepsilon_{kl,m} \delta \varepsilon_{ij} dV + \int \Pi_{ijmkln,n} \delta \varepsilon_{ij} \delta \varepsilon_{kl} \hat{n}_m dS. \tag{32}$$

Later in Part 5, we find that $\Pi_{ijmkln,n} \delta \varepsilon_{ij,m} \delta \varepsilon_{kl} = \Pi_{ijmkln,n} \delta \varepsilon_{kl,m} \delta \varepsilon_{ij}$ is valid for any solids that could be described by EAM-type potentials. Thus, the second integral in the last equality of (32) is equal to the left hand side of equation (32), which results in

$$\int \Pi_{ijmkln,n} \delta \varepsilon_{ij,m} \delta \varepsilon_{kl} dV = -\frac{1}{2} \int \Pi_{ijmkln,mn} \delta \varepsilon_{kl} \delta \varepsilon_{ij} dV + \frac{1}{2} \int \Pi_{ijmkln,n} \delta \varepsilon_{ij} \delta \varepsilon_{kl} \hat{n}_m dS. \tag{33}$$

Substituting the above equation into (31) and further substituting the obtained result into (30), we obtain

$$\delta^2 \bar{G} = \int \left( C_{ijkl} - W_{ijkln,n} + \frac{1}{2} \Pi_{ijmkln,mn} \right) \delta \varepsilon_{ij} \delta \varepsilon_{kl} dV - \int \Pi_{ijmkln} \delta \varepsilon_{ij,mn} \delta \varepsilon_{kl} dV$$

$$+ \int \Pi_{ijmkln} \delta \varepsilon_{ij,m} \delta \varepsilon_{kl} \hat{n}_n dS - \frac{1}{2} \int \Pi_{ijmkln,n} \delta \varepsilon_{ij} \delta \varepsilon_{kl} \hat{n}_m dS > 0. \tag{34}$$

Under the fixed boundary condition, all surface contributions in (34) vanish. And the higher order derivatives of strain gradients, representing higher order effects of the strain gradients, are omitted in present work. Then the stability condition (34) could be expressed as

$$\left(C_{ijkl} - W_{ijkln,n} + \frac{1}{2}\Pi_{ijmkln,mn}\right)\delta\varepsilon_{ij}\delta\varepsilon_{kl} > 0, \tag{35}$$

which should be satisfied for arbitrary $\delta\boldsymbol{\varepsilon}$. If we define

$$\widetilde{K}_{ijkl} = C_{ijkl} - W_{ijkln,n} + \frac{1}{2}\Pi_{ijmkln,mn}, \tag{36}$$

condition (35) requires that $\widetilde{\boldsymbol{K}}$ are positive definite. The condition has considered contributions of both strain and strain gradient, which is able to judge mechanical stabilities of solids bearing either strain, strain gradient or their combinations. Besides, the equilibrium equation (19b) becomes

$$t_{ij} = \sigma_{ij} - \tau_{ijm,m}. \tag{37}$$

This equation establishes relationships between external stresses and the work-conjugates to strains and strain gradients. Generally, $\boldsymbol{\Pi}$ (as well as other conjugate variables) is a function of strain. Without considering the higher order effects of the strain gradients, we have

$$\Pi_{ijmkln,mn} = \frac{\partial^2 \Pi_{ijmkln}}{\partial \varepsilon_{rt} \partial \varepsilon_{op}} \varepsilon_{rt,m} \varepsilon_{rt,n}. \tag{38}$$

If no strain gradients are present, then $\tau_{ijm}$ and $W_{ijkln}$ are zeros (Wang et al., 2017). Obviously, $\Pi_{ijmkln,mn}$ are also zeros for any combinations of (*ijkl*) at zero strain gradients. Thus, the equilibrium equation and stability condition will reduce to

$$t_{ij} = \sigma_{ij}, \tag{39}$$

and

$$C_{ijkl}\delta\varepsilon_{ij}\delta\varepsilon_{kl} > 0, \tag{40}$$

respectively, which is consistent with the Born stability conditions. Then equation (39) indicates that the external stress is balanced by internal stress —— the work-conjugate to small strains, and thereby $\boldsymbol{\sigma}$ represents the Cauchy (true) stress. In the next section, the stability condition (35) will be generalized for finite strains.

## 4. Strain-Gradient Related Stability Criteria for Finite Strains

Under dynamic loadings, deformations are usually not small before the mechanical instabilities take place. Thus, the above theory should be reformulated at finite strains. Supposing that a solid with an initial configuration $\{\boldsymbol{a}\}$ is deformed in current configuration $\{\boldsymbol{X}\}$ under a Lagrangian finite strain $\boldsymbol{\eta}$, then instabilities of the solid at $\{\boldsymbol{X}\}$ could be checked through exerting a small virtual strain ($\boldsymbol{\varepsilon}$), as well as virtual strain gradient ($\nabla\boldsymbol{\varepsilon}$) disturbances. The resulting configuration is supposed to be $\{\boldsymbol{Y}\}$. Let $\alpha_{ik} = \frac{\partial X_i}{\partial a_k}$ and $\bar{\alpha}_{ik} = \frac{\partial Y_i}{\partial X_k}$ to be the deformation gradient tensors associated with $\boldsymbol{\eta}$ and $\boldsymbol{\varepsilon}$, respectively. According to definitions of small linear strain as and Lagrangian finite strain, we have

$$\eta_{ij} = \frac{1}{2}(\alpha_{ki}\alpha_{kj} - \delta_{ij}), \tag{41}$$

$$\varepsilon_{ij} = \bar{\alpha}_{ij} - \delta_{ij}, \tag{42}$$

where $\delta_{ij}$ is the Kronecker delta. Further, $\boldsymbol{\varepsilon}$ measured in $\{\boldsymbol{a}\}$ is denoted by $\boldsymbol{e}$ whose deformation gradient tensor is $\hat{\boldsymbol{\alpha}}$. Then, we have

$$\hat{\alpha}_{ik} = \bar{\alpha}_{ij}\alpha_{jk} = \alpha_{ik} + \varepsilon_{ij}\alpha_{jk}, \tag{43}$$

$$e_{ij} = \frac{1}{2}(\hat{\alpha}_{ki}\hat{\alpha}_{kj} - \delta_{ij}). \tag{44}$$

Using (43), as well as (41), to rewriting the above equation, we get

$$e_{ij} = \eta_{ij} + \frac{1}{2}\varepsilon_{kl}(\alpha_{ki}\alpha_{lj} + \alpha_{kj}\alpha_{li}) + \frac{1}{2}\alpha_{li}\alpha_{mj}\varepsilon_{kl}\varepsilon_{km}. \tag{45}$$

Since the linear strain is symmetric, this leads to relations of $\varepsilon_{kl}\alpha_{ki}\alpha_{lj} = \varepsilon_{kl}\alpha_{kj}\alpha_{li}$. And the second order terms of $\boldsymbol{\varepsilon}$ are omitted because $\boldsymbol{\varepsilon}$ is a small quantity. Thus, equation (45) is reduced to

$$e_{ij} = \eta_{ij} + \varepsilon_{kl}\alpha_{ki}\alpha_{lj}, \tag{46}$$

which is consistent with relations given by Tröster and Schranz (Tröster et al., 2002). The virtual strain gradients in $\{\boldsymbol{a}\}$ and $\{\boldsymbol{X}\}$ are defined by

$$\eta_{ij,m} = \frac{\partial \eta_{ij}}{\partial a_m}, \tag{47}$$

$$e_{ij,m} = \frac{\partial e_{ij}}{\partial a_m}, \tag{48}$$

$$\varepsilon_{ij,M} = \frac{\partial \varepsilon_{ij}}{\partial X_M}. \tag{49}$$

According to equation (45), the above two strain gradients are related by

$$e_{ij,m} = \eta_{ij,m} + \alpha_{Mm}\alpha_{ki}\alpha_{lj}\varepsilon_{kl,M} + \frac{1}{2}\alpha_{li}\alpha_{nj}\alpha_{Mm}(\varepsilon_{kl,M}\varepsilon_{kn} + \varepsilon_{kl}\varepsilon_{kn,M}). \tag{50}$$

From (46) and (50), the variation of $\boldsymbol{\eta}$ and $\nabla\boldsymbol{\eta}$ with respect to the virtual strain ($\boldsymbol{\varepsilon}$) and strain gradient ($\nabla\boldsymbol{\varepsilon}$) could be written as

$$\delta\eta_{ij} = e_{ij} - \eta_{ij} = \alpha_{ki}\alpha_{lj}\varepsilon_{kl}, \tag{51}$$

$$\delta\eta_{ij,m} = e_{ij,m} - \eta_{ij,m} = \alpha_{ki}\alpha_{lj}\alpha_{Mm}\varepsilon_{kl,M} + \frac{1}{2}(\alpha_{li}\alpha_{nj} + \alpha_{ni}\alpha_{lj})\alpha_{Mm}\varepsilon_{kl,M}\varepsilon_{kn}. \tag{52}$$

For the same reasons as the expression (24), the Helmholtz free energy could be expressed as

$$\rho_a \bar{F}(\boldsymbol{\eta}, \nabla\boldsymbol{\eta}, T) = -\rho_a \bar{F}(\boldsymbol{0}, \boldsymbol{0}, T) + \hat{\sigma}_{ij}\eta_{ij} + \hat{\tau}_{ijm}\eta_{ij,m} + \frac{1}{2}\hat{C}_{ijkl}\eta_{ij}\eta_{kl} + \widehat{W}_{ijkln}\eta_{ij}\eta_{kl,n}$$

$$+ \frac{1}{2}\widehat{\Pi}_{ijmkln}\eta_{ij,m}\eta_{kl,n}, \tag{53}$$

where $\rho_a = 1/V(\boldsymbol{a})$, work-conjugates to strain or strain gradient in above equation are defined by

$$\hat{\sigma}_{ij} = \frac{1}{V(\boldsymbol{a})}\frac{\partial \bar{F}}{\partial \eta_{ij}}, \tag{54}$$

$$\hat{C}_{ijkl} = \frac{1}{V(\boldsymbol{a})}\frac{\partial^2 \bar{F}}{\partial \eta_{ij}\partial \eta_{kl}}, \tag{55}$$

$$\hat{\tau}_{ijm} = \frac{1}{V(\boldsymbol{a})}\frac{\partial \bar{F}}{\partial \eta_{ij,m}}, \tag{56}$$

$$\widehat{W}_{ijkln} = \frac{1}{V(\boldsymbol{a})}\frac{\partial^2 \bar{F}}{\partial \eta_{ij}\partial \eta_{kl,n}}, \tag{57}$$

$$\widehat{\Pi}_{ijmkln} = \frac{1}{V(\boldsymbol{a})}\frac{\partial^2 \bar{F}}{\partial \eta_{ij,m}\partial \eta_{kl,n}}. \tag{58}$$

In the following, we will reformulate the equilibrium equation and the stability condition for finite strains. According to (51)-(53), variations of the free energy with respect to the small virtual strain ($\boldsymbol{\varepsilon}$) is

$$\delta\bar{F} = \int \delta f(\boldsymbol{\eta}, \nabla\boldsymbol{\eta}, \boldsymbol{a})dV_a$$

$$= \int (\hat{\sigma}_{ij}\delta\eta_{ij} + \hat{\tau}_{ijm}\delta\eta_{ij,m} + \hat{C}_{ijkl}\eta_{kl}\delta\eta_{ij} + \widehat{W}_{ijkln}\delta\eta_{ij}\eta_{kl,n} + \widehat{W}_{ijkln}\eta_{ij}\delta\eta_{kl,n}$$

$$+\widehat{\Pi}_{ijmkln}\eta_{ij,m}\delta\eta_{kl,n})\,dV_a$$
$$=\int[(\hat{\sigma}_{ij}+\hat{C}_{ijkl}\eta_{kl}+\widehat{W}_{ijkln}\eta_{kl,n})\delta\eta_{ij}$$
$$+(\hat{\tau}_{kln}+\widehat{W}_{ijkln}\eta_{ij}+\widehat{\Pi}_{ijmkln}\eta_{ij,m})\delta\eta_{kl,n}]\det([\boldsymbol{\alpha}]^{-1})dV_{\mathbf{X}}. \tag{59}$$

Assuming that $\boldsymbol{\alpha}$ is symmetric (only a rotational part is omitted since the rotation does not modify physical properties of solids), it could be uniquely determined by $\boldsymbol{\eta}$ (See Eq. 41). Then we have expansions of $\boldsymbol{\alpha}$ in terms of $\boldsymbol{\eta}$, that is,

$$\alpha_{ij}=\delta_{ij}+\eta_{ij}-\frac{1}{2}\eta_{ik}\eta_{kj}+\cdots. \tag{60}$$

With the above expansion, (51) and (52) could be expressed as

$$\delta\eta_{ij}=\alpha_{Ii}\alpha_{Jj}\varepsilon_{IJ}=\left(\delta_{Ii}+\eta_{Ii}-\frac{1}{2}\eta_{IM}\eta_{Mi}+\cdots\right)\left(\delta_{Jj}+\eta_{Jj}-\frac{1}{2}\eta_{JN}\eta_{Nj}+\cdots\right)\varepsilon_{IJ}$$
$$=\left(\delta_{Ii}\delta_{Jj}+\delta_{Ii}\eta_{Ji}+\delta_{Jj}\eta_{Ii}+\eta_{Ii}\eta_{Jj}-\frac{1}{2}\delta_{Ii}\eta_{JN}\eta_{Nj}-\frac{1}{2}\delta_{Jj}\eta_{IM}\eta_{Mi}+\cdots\right)\varepsilon_{IJ}, \tag{61}$$

and

$$\delta\eta_{kl,n}=\alpha_{Kk}\alpha_{Ll}\alpha_{Nn}\varepsilon_{KL,N}$$
$$=(\delta_{Kk}+\eta_{Kk}-\cdots)(\delta_{Ll}+\eta_{Ll}-\cdots)(\delta_{Nn}+\eta_{Nn}-\cdots)\varepsilon_{KL,N}$$
$$=(\delta_{Kk}\delta_{Ll}\delta_{Nn}+\delta_{Kk}\delta_{Ll}\eta_{Nn}+\delta_{Kk}\eta_{Ll}\delta_{Nn}+\eta_{Kk}\delta_{Ll}\delta_{Nn}+\cdots)\varepsilon_{KL,N}. \tag{62}$$

Substituting (61), (62) and the below relation, i.e.,

$$\det([\boldsymbol{\alpha}]^{-1})\approx 1-\eta_{ii}, \tag{63}$$

into (59) and rearranging the obtained terms (See Appendix B), we get

$$\delta\bar{F}=\int[\hat{\sigma}_{IJ}-\hat{\tau}_{IJN,N}+(\hat{B}_{IJkl}+\hat{\tau}_{IJl,k}-\widehat{\Lambda}_{klIJN,N})\eta_{kl}+(\widehat{W}_{IJkln}-\widehat{\Lambda}_{klIJn}-\widehat{\Pi}_{klnIJN,N})\eta_{kl,n}+$$
$$\widehat{\Pi}_{ijmIJN}\eta_{ij,mN}]\varepsilon_{IJ}\,dV_{\mathbf{X}}+\int(\hat{\tau}_{KLN}+\widehat{\Lambda}_{ijKLN}\eta_{ij}+\widehat{\Pi}_{ijmKLN}\eta_{ij,m})\varepsilon_{KL}\hat{n}_N\,dS, \tag{64}$$

where $\mathbf{n}$ is outward unit vector normal to surface ($S$) of $V_{\mathbf{X}}$ and

$$\hat{B}_{IJkl}=\hat{C}_{IJkl}+\hat{\sigma}_{Il}\delta_{kJ}+\hat{\sigma}_{IJ}\delta_{kI}-\hat{\sigma}_{IJ}\delta_{kl}, \tag{65}$$
$$\widehat{\Lambda}_{ijKLN}=\widehat{W}_{ijKLN}+\hat{\tau}_{KLj}\delta_{iN}+\hat{\tau}_{KjN}\delta_{iL}+\hat{\tau}_{jLN}\delta_{iK}-\hat{\tau}_{KLN}\delta_{ij}. \tag{66}$$

Considering the expression of $\delta W$ (See Eq.16) and neglecting effects of higher order strain gradients (i.e., terms containing $\eta_{ij,mN}$ in the Eq.64), the equilibrium condition requires

$$\delta\bar{F}-\delta\bar{W}=\delta\bar{F}-\int t_{IJ}\varepsilon_{IJ}dV_{\mathbf{X}}$$
$$=\int[\hat{\sigma}_{IJ}-\hat{\tau}_{IJN,N}-t_{IJ}+(\hat{B}_{IJkl}+\hat{\tau}_{IJl,k}-\widehat{\Lambda}_{klIJN,N})\eta_{kl}$$
$$+(\widehat{W}_{IJkln}-\widehat{\Lambda}_{klIJn}-\widehat{\Pi}_{klnIJN,N})\eta_{kl,n}]\varepsilon_{IJ}\,dV_{\mathbf{X}}$$
$$+\int(\hat{\tau}_{KLN}+\widehat{\Lambda}_{ijKLN}\eta_{ij}+\widehat{\Pi}_{ijmKLN}\eta_{ij,m})\varepsilon_{KL}\hat{n}_N\,dS=0$$
$$\tag{67}$$

To make the last equality in (67) satisfied for arbitrary $\boldsymbol{\varepsilon}$, the expression in the square bracket of the first integral should be zero, that is

$$t_{ij}=\hat{\sigma}_{ij}-\hat{\tau}_{ijm,m}+(\hat{B}_{ijkl}+\hat{\tau}_{IJl,k}-\widehat{\Lambda}_{klijN,N})\eta_{kl}+(\widehat{W}_{ijkln}-\widehat{\Lambda}_{klijn}-\widehat{\Pi}_{klnijN,N})\eta_{kl,n}. \tag{68a}$$

It is worth noting that $\hat{\boldsymbol{\sigma}}$ and $\hat{\boldsymbol{\tau}}$ (as the other elastic constants) are measured at initial configuration, i.e., $\{\boldsymbol{a}\}$, while $\mathbf{t}$ is the extra stress balanced by the internal stresses of configuration $\{\boldsymbol{X}\}$. A special, but important, case is that the initial configuration is free of strain or strain gradient. In such case, we have $\hat{\sigma}_{ij}-\hat{\tau}_{ijm,m}=0$ because of Eq. (37). Then the equilibrium equation is simplified to be

$$t_{ij}=(\hat{B}_{ijkl}+\hat{\tau}_{IJl,k}-\widehat{\Lambda}_{klijN,N})\eta_{kl}+(\widehat{W}_{ijkln}-\widehat{\Lambda}_{klijn}-\widehat{\Pi}_{klnijN,N})\eta_{kl,n}, \tag{68b}$$

which indicates that internal stresses are linearly changed with strain and strain gradients. For centrosymmetric crystals where $\hat{\boldsymbol{\tau}}$ and $\widehat{\boldsymbol{W}}$ are zeros, Eq. (68b) becomes

$$t_{ij} = \hat{C}_{ijkl}\eta_{kl} - \widehat{\Pi}_{klnijN,N}\eta_{kl,n}. \tag{68c}$$

If the crystals are free of strain gradient, Eq. (68c) will reduce to Hooke's law. Specially, for small $\boldsymbol{\eta}$, Eq. (68b) or (68c) should be equivalent to Eq. (37), which results in

$$\hat{\sigma}_{ij}(\mathbf{X}) = \left(\hat{B}_{ijkl} + \hat{\tau}_{ijl,k} - \widehat{\Lambda}_{klijN,N}\right)\eta_{kl} \equiv \hat{C}_{ijkl}(\mathbf{X})\eta_{kl}, \tag{69a}$$

$$\hat{\tau}_{ijm,m}(\mathbf{X}) = -\left(\widehat{W}_{ijkln} - \widehat{\Lambda}_{klijn} - \widehat{\Pi}_{klnijN,N}\right)\eta_{kl,n} \equiv \widehat{\Pi}_{klnijm,m}(\mathbf{X})\eta_{kl,n}. \tag{69b}$$

or

$$\hat{\sigma}_{ij}(\mathbf{X}) = \hat{C}_{ijkl}\eta_{kl}, \quad \hat{\tau}_{ijm,m}(\mathbf{X}) = \widehat{\Pi}_{klnijm,m}\eta_{kl,n}. \tag{70}$$

In Eq. (69a) and (69b), the elastic constants at configuration $\{X\}$ relate to the ones at $\{a\}$ by

$$\hat{C}_{ijkl}(\mathbf{X}) = \hat{B}_{ijkl} + \hat{\tau}_{ijl,k} - \widehat{\Lambda}_{klijn,n}, \quad \widehat{\Pi}_{klnijm,m}(\mathbf{X}) = -\widehat{W}_{ijkln} + \widehat{\Lambda}_{klijn} + \widehat{\Pi}_{klnijn,n}. \tag{71}$$

Thus, the equilibrium equation (68b) or (68c) could expressed in configuration $\{X\}$ by

$$t_{ij} = \hat{\sigma}_{ij}(\mathbf{X}) - \hat{\tau}_{ijm,m}(\mathbf{X}). \tag{68d}$$

In contrast to the strain-related elastic constitutive relation (69a), Eq. (69b) represents strain-gradient related elastic constitutive relation. The stability condition for finite strains requires

$$\delta^2 \bar{F} = \int \left[\left(\hat{B}_{IJkl} + \hat{\tau}_{IJl,k} - \widehat{\Lambda}_{klIJN,N}\right)\delta\eta_{kl} + \left(\widehat{W}_{IJkln} - \widehat{\Lambda}_{klIJn} - \widehat{\Pi}_{klnIJN,N}\right)\delta\eta_{kl,n}\right]\varepsilon_{IJ}\,\mathrm{d}V_{\mathbf{X}} > 0, \tag{72}$$

where the surface contribution is omitted since the surface only serves as a virtual bound used to distinguish the inner (materials of interest) for the outer. Furthermore, only bulk properties of solid are concerned here. Substituting (61) and (62) into above expression and retaining up to the second order small quantities of strains and strain gradients, we have

$$\delta^2 \bar{F} = \int \left[\left(\hat{B}_{IJKL} + \hat{\tau}_{IJL,K} - \widehat{\Lambda}_{KLIJN,N}\right)\varepsilon_{KL} + \left(\widehat{W}_{IJKLM} - \widehat{\Lambda}_{KLIJM} - \widehat{\Pi}_{KLMIJN,N}\right)\varepsilon_{KL,M}\right]\varepsilon_{IJ}\mathrm{d}V_{\mathbf{X}}$$

$$= \int\left(\hat{B}_{IJKL} + \hat{\tau}_{IJL,K} - \widehat{\Lambda}_{KLIJN,N}\right)\varepsilon_{IJ}\,\varepsilon_{KL}\mathrm{d}V_{\mathbf{X}} + \int\left(\widehat{W}_{IJKLM} - \widehat{\Lambda}_{KLIJM} - \widehat{\Pi}_{KLMIJN,N}\right)\varepsilon_{KL,M}\varepsilon_{IJ}\mathrm{d}V_{\mathbf{X}}. \tag{73}$$

From above equation, we could conclude that not only strain (represented by $\widehat{\mathbf{B}}$), but also strain gradient (represented by $\hat{\boldsymbol{\tau}}$, $\widehat{\mathbf{W}}$ (or $\widehat{\boldsymbol{\Lambda}}$) and $\widehat{\boldsymbol{\Pi}}$), govern the mechanical stabilities of solids. And the $\hat{\boldsymbol{\tau}}$ and $\widehat{\mathbf{W}}$ (or $\widehat{\boldsymbol{\Lambda}}$) are nearly zero and thus be omitted for small strain gradients, contribution of the strain gradient on the mechanical stabilities mainly result from the second integral in Eq. (73) through $\widehat{\boldsymbol{\Pi}}$. Obviously, if $-\widehat{\Pi}_{KLMIJN,N}$ is positive definite, then result of the second integral will be positive. In other words, solids are stable under disturbances of strain gradients if $\hat{T}_{KLMIJN,N}$ is negative definite. This conclusion has been justified in single crystal iron under ramp compressions with various strain rates. Because the Eq. (73) only reflects the separate effects of strain and strain gradient, respectively, total effects on the stabilities are still unknown. To make clear of the total effects, the second integral in the Eq. (73) is further simplified as follows:

$$\int\left(\widehat{W}_{IJKLM} - \widehat{\Lambda}_{KLIJM} - \widehat{\Pi}_{KLMIJN,N}\right)\varepsilon_{KL,M}\varepsilon_{IJ}\mathrm{d}V_{\mathbf{X}}$$

$$= -\int\left[\left(\widehat{W}_{IJKLM} - \widehat{\Lambda}_{KLIJM} - \widehat{\Pi}_{KLMIJN,N}\right)\varepsilon_{IJ}\right]_{,M}\varepsilon_{KL}\mathrm{d}V_{\mathbf{X}}$$

$$\quad + \int\left(\widehat{W}_{IJKLM} - \widehat{\Lambda}_{KLIJM} - \widehat{\Pi}_{KLMIJN,N}\right)\varepsilon_{KL}\varepsilon_{IJ}\hat{n}_M\mathrm{d}S$$

$$= -\int\left(\widehat{W}_{IJKLM} - \widehat{\Lambda}_{KLIJM} - \widehat{\Pi}_{KLMIJN,N}\right)\varepsilon_{IJ,M}\varepsilon_{KL}\mathrm{d}V_{\mathbf{X}}$$

$$\quad - \int\left(\widehat{W}_{IJKLM} - \widehat{\Lambda}_{KLIJM} - \widehat{\Pi}_{KLMIJN,N}\right)_{,M}\varepsilon_{IJ}\varepsilon_{KL}\mathrm{d}V_{\mathbf{X}}$$

$$\quad + \int\left(\widehat{W}_{IJKLM} - \widehat{\Lambda}_{KLIJM} - \widehat{\Pi}_{KLMIJN,N}\right)\varepsilon_{KL}\varepsilon_{IJ}\hat{n}_M\mathrm{d}S$$

$$= -\int\left(\widehat{W}_{KLIJM} - \widehat{\Lambda}_{IJKLM} - \widehat{\Pi}_{IJMKLN,N}\right)\varepsilon_{KL,M}\varepsilon_{IJ}\mathrm{d}V_{\mathbf{X}}$$

$$-\int \left(\widehat{W}_{IJKLM} - \widehat{\Lambda}_{KLIJM} - \widehat{\Pi}_{KLMIJN,N}\right)_{,M} \varepsilon_{IJ} \varepsilon_{KL} dV_{\mathbf{X}}$$

$$+ \int \left(\widehat{W}_{IJKLM} - \widehat{\Lambda}_{KLIJM} - \widehat{\Pi}_{KLMIJN,N}\right) \varepsilon_{KL} \varepsilon_{IJ} \hat{n}_M dS$$

With aids of the equality below

$$\left(\widehat{W}_{IJKLM} - \widehat{\Lambda}_{KLIJM} - \widehat{\Pi}_{KLMIJN,N}\right) \varepsilon_{KL,M} \varepsilon_{IJ} = \left(\widehat{W}_{KLIJM} - \widehat{\Lambda}_{IJKLM} - \widehat{\Pi}_{IJMKLN,N}\right) \varepsilon_{KL,M} \varepsilon_{IJ}, \quad (74)$$

we obtain

$$\int \left(\widehat{W}_{IJKLM} - \widehat{\Lambda}_{KLIJM} - \widehat{\Pi}_{KLMIJN,N}\right) \varepsilon_{KL,M} \varepsilon_{IJ} dV_{\mathbf{X}}$$

$$= -\frac{1}{2} \int \left(\widehat{W}_{IJKLM} - \widehat{\Lambda}_{KLIJM} - \widehat{\Pi}_{KLMIJN,N}\right)_{,M} \varepsilon_{IJ} \varepsilon_{KL} dV_{\mathbf{X}}$$

$$+ \frac{1}{2} \int \left(\widehat{W}_{IJKLM} - \widehat{\Lambda}_{KLIJM} - \widehat{\Pi}_{KLMIJN,N}\right) \varepsilon_{KL} \varepsilon_{IJ} \hat{n}_M dS. \quad (75)$$

Finally, expression (72) reduces to

$$\delta^2 F = \int \left(\widehat{B}_{IJKL} + \hat{\tau}_{IJL,K} - \widehat{\Lambda}_{KLIJN,N}\right) \varepsilon_{IJ} \varepsilon_{KL} dV_{\mathbf{X}}$$

$$-\frac{1}{2} \int \left(\widehat{W}_{IJKLM} - \widehat{\Lambda}_{KLIJM} - \widehat{\Pi}_{KLMIJN,N}\right)_{,M} \varepsilon_{IJ} \varepsilon_{KL} dV_{\mathbf{X}}$$

$$+ \frac{1}{2} \int \left(\widehat{W}_{IJKLM} - \widehat{\Lambda}_{KLIJM} - \widehat{\Pi}_{KLMIJN,N}\right) \varepsilon_{KL} \varepsilon_{IJ} \hat{n}_M dS$$

$$= \int \left(\widehat{B}_{IJKL} + \hat{\tau}_{IJL,K} - \frac{1}{2}\widehat{W}_{IJKLM,M} - \frac{1}{2}\widehat{\Lambda}_{KLIJM,M} + \frac{1}{2}\widehat{\Pi}_{KLMIJN,NM}\right) \varepsilon_{IJ} \varepsilon_{KL} dV_{\mathbf{X}}$$

$$+ \frac{1}{2} \int \left(\widehat{W}_{IJKLM} - \widehat{\Lambda}_{KLIJM} - \widehat{\Pi}_{KLMIJN,N}\right) \varepsilon_{KL} \varepsilon_{IJ} \hat{n}_M dS$$

$$= \int \left(\widehat{B}_{IJKL} + \hat{\tau}_{IJL,K} - \frac{1}{2}\widehat{W}_{IJKLM,M} - \frac{1}{2}\left(\widehat{W}_{KLIJM,M} + \hat{\tau}_{IJL,M}\delta_{KM} + \hat{\tau}_{ILM,M}\delta_{KJ} + \hat{\tau}_{LJM,M}\delta_{KI} -\right.\right.$$

$$\left.\left.\hat{\tau}_{IJM,M}\delta_{KL}\right) + \frac{1}{2}\widehat{\Pi}_{KLMIJN,NM}\right) \varepsilon_{IJ} \varepsilon_{KL} dV_{\mathbf{X}} + \frac{1}{2} \int \left(\widehat{W}_{IJKLM} - \widehat{\Lambda}_{KLIJM} - \widehat{\Pi}_{KLMIJN,N}\right) \varepsilon_{KL} \varepsilon_{IJ} \hat{n}_M dS$$

$$= \int \left(\widehat{B}_{IJKL} + \hat{\tau}_{IJL,K} - \widehat{W}_{IJKLM,M} - \frac{1}{2}\left(\hat{\tau}_{IJL,M}\delta_{KM} + \hat{\tau}_{ILM,M}\delta_{KJ} + \hat{\tau}_{LJM,M}\delta_{KI} - \hat{\tau}_{IJM,M}\delta_{KL} - \right.\right.$$

$$\left.\left.\widehat{\Pi}_{KLMIJN,NM}\right)\right) \varepsilon_{IJ} \varepsilon_{KL} dV_{\mathbf{X}} + \frac{1}{2} \int \left(\widehat{W}_{IJKLM} - \widehat{\Lambda}_{KLIJM} - \widehat{\Pi}_{KLMIJN,N}\right) \varepsilon_{KL} \varepsilon_{IJ} \hat{n}_M dS$$

$$= \int \widehat{K}_{IJKL} \varepsilon_{IJ} \varepsilon_{KL} dV_{\mathbf{X}} + \frac{1}{2} \int \left(\widehat{W}_{IJKLM} - \widehat{\Lambda}_{KLIJM} - \widehat{\Pi}_{KLMIJN,N}\right) \varepsilon_{KL} \varepsilon_{IJ} \hat{n}_M dS > 0, \quad (76)$$

where

$$\widehat{K}_{IJKL} = \widehat{B}_{IJKL} + \widehat{\Theta}_{IJKL}, \quad (77)$$

$$\widehat{\Theta}_{IJKL} = \hat{\tau}_{IJL,K} - \widehat{W}_{IJKLM,M}$$

$$- \frac{1}{2}\left(\hat{\tau}_{IJL,M}\delta_{KM} + \hat{\tau}_{ILM,M}\delta_{KJ} + \hat{\tau}_{LJM,M}\delta_{KI} - \hat{\tau}_{IJM,M}\delta_{KL} - \widehat{\Pi}_{IJNKLM,MN}\right). \quad (78)$$

In the derivation before the last line of (76), we have used the equality of $\widehat{W}_{IJKLM,M}\varepsilon_{IJ}\varepsilon_{KL} = \widehat{W}_{KLIJM,M}\varepsilon_{IJ}\varepsilon_{KL}$. Again, the surface contribution vanishes in the last inequality of expression (76). And, to satisfy condition (76) for arbitrary small virtual strain $\boldsymbol{\varepsilon}$, $\widehat{\mathbf{K}}$ should be positive definite. Generally speaking, $\widehat{\mathbf{K}}$ ($\widehat{\mathbf{B}}$ and $\widehat{\mathbf{G}}$) are not symmetric. For centrosymmetric crystals, it can be shown that $\hat{\boldsymbol{\tau}}$ and $\widehat{\mathbf{W}}$ are zeros for small $\nabla\boldsymbol{\eta}$. Thereby, $\widehat{\boldsymbol{\Theta}}$ reduces to

$$\widehat{\Theta}_{IJKL} = \frac{1}{2}\widehat{\Pi}_{IJNKLM,MN}, \quad (79)$$

where

$$\widehat{\Pi}_{IJNKLM,MN} = \frac{\partial^2 \widehat{\Pi}_{IJNKLM}}{\partial \eta_{RT} \partial \eta_{OP}} \eta_{RT,M} \eta_{OP,N} \tag{80}$$

due to the same reasons as explained for the expression (38). If the strain gradient is not present, then $\widehat{K}_{IJKL}$ is equal the *Birch coefficients* ($\widehat{B}_{IJKL}$), which is usually expressed in a symmetric form with respect to the exchange of $(IJ)$ and $(KL)$. In this case, the stability condition established here is consistent with the *B criteria* (Wang et al., 1995; Wang et al., 1993). Besides, from expression (80), we find that $\widehat{T}_{IJNKLM,MN}$ is symmetric with respect to $(IJ)\leftrightarrow(KL)$ since $(IJN)$ is exchangeable with $(KLM)$ for $\widehat{T}_{IJNKLM}$ and $M$ and $N$ are exchangeable for $\widehat{\Pi}_{IJNKLM,MN}$. Thus, $\widehat{K}$ poses $(IJ)\leftrightarrow(KL)$ symmetry. With Voigt notation, $\widehat{K}$ could be expressed as a 6-D symmetry matrix which has six real eigenvalues. Finally, the stability condition is equivalent to

$$\widehat{K}_{min} = \text{MIN}\{\widehat{K}_\mu^{eig} | \mu = 1,2,\cdots,6\} > 0, \tag{81}$$

where $\widehat{K}_\mu^{eig}$ denote the $\mu$-th ($\mu$ = 1, 2, ..., 6) eigenvalue of $\widehat{K}$. It needs to point out that the stability condition is irrespective to **η** because only the second order small quantities are retained in our derivations. This means that the stability condition is valid when the finite strain (**η**) of configuration {**X**} is not very large. The stability condition for finite strains should be equivalent to the one for small strains when we take configuration {**X**} as the initial configuration, i.e., **η = 0**. Alternatively, through replacing the elastic constants at the initial configure in Eq. (36) by the ones at {**X**} according to relations (71), one should obtain Eq. (77). To fulfill this statement, we find that the divergence of $\widehat{W}$ in Eq. (36) must be zero, i.e.,

$$\widehat{W}_{IJKLM,M}(\boldsymbol{X}) = 0. \tag{82}$$

Our recent work (Wang et al., 2017) shows that $\widehat{W}$ is zero for centro-symmetric crystals, which satisfies the above equation. The results shown in Eq. (82) further suggests that the divergence of $\widehat{W}$ is zero for arbitrary solids. Perhaps, this is the results of the second order approximation of strain and strain gradient. We will show in present work that the results obtained under the approximation are valid for various metals under a strain rate up to more than $10^{11}$ s$^{-1}$.

## 5. Microscopic Expressions for the Work-Conjugates to Strain and Strain Gradient

Application of the strain-gradient related stability criteria to real solids requires to known the correct strain-gradient elastic constants of the corresponding solids. In contrast to the indirection approaches developed for estimation of the strain-gradient elastic constants (DiVincenzo, 1986; Maranganti and Sharma, 2007; Stengel, 2013, 2016), we present direct microscopic expressions for the elastic constants in a static manner, which enable us to evaluate the higher order elastic stabilities of real solids conveniently without the need to assume that the solid is linear isotropic. In this part, we use lowercase Greek letters, such as $\alpha$, $\beta$, $\gamma$, $\mu$, $\nu$, $\lambda$ and $\rho$, to distinguish the three Cartesian components of vectors or tensors, and lowercase English letters, such as $i$ and $j$, to stand for atom indexes. Summation over repeated indexes is only applied for the Cartesian indexes.

Imaging that a small spherical volume (*V*) in a solid, consisting of *N* atoms, is deformed from reference configuration {**X**} to current configuration {**Y**} under a uniform strain gradient (∇**η**), we will evaluate all elastic constants as defined by Eq. (25)-(29) for the volume. Without loss of generality, we take the center of the volume as origin point of position vector. Strain at the origin point is assume to be zero, which means that deformation is always measured with respect to states at the center of the volume. Thus, our calculated elastic constants will change with the position of the volume center. This is reasonable since strains are not uniform at presence of the strain gradient. Pairwise separation between atom $i$ and $j$ in configuration $\{X\}$ and $\{Y\}$ are denoted by $\mathbf{R}_{ij} = \mathbf{X}_i - \mathbf{X}_j$ and $\mathbf{r}_{ij} = \mathbf{Y}_i - \mathbf{Y}_j$, respectively. Since $\mathbf{X}_0 = \mathbf{Y}_0 = \mathbf{0}$, we will write $\mathbf{r}_{i0}$ ($\mathbf{R}_{i0}$) as $\mathbf{r}_i$ ($\mathbf{R}_i$). Then, displacement of atom $i$ at $\mathbf{X}_i$ could be expanded nearby the origin point, to the second order, in terms of the position vector, that is

$$u_\alpha(\mathbf{X}_i) = u_\alpha(0) + u_{\alpha,\beta}(0) X_i^\beta + \frac{1}{2} u_{\alpha,\beta\gamma}(0) X_i^\beta X_i^\gamma, \tag{83}$$

where $u_\alpha(\mathbf{X}_i) = Y_i^\alpha - X_i^\alpha$ ($\alpha$, $\beta$, $\gamma$ = 1,2,3), $u_{\alpha,\beta}$ and $u_{\alpha,\beta\gamma}$ are the first and second gradient of the displacement. This expression is precise enough for metals since effective interactions between atoms are short-ranged. The Lagrangian strain and strain gradient relate to the gradients of displacement by

$$\eta_{\alpha\beta} = \frac{1}{2}(u_{\alpha,\beta} + u_{\beta,\alpha} + u_{\mu,\alpha} u_{\mu,\beta}), \tag{84}$$

$$\eta_{\alpha\beta,\gamma} = \frac{1}{2}(u_{\alpha,\beta\gamma} + u_{\beta,\alpha\gamma} + u_{\mu,\alpha\gamma} u_{\mu,\beta} + u_{\mu,\alpha} u_{\mu,\beta\gamma}). \tag{85}$$

Average energy density of the deformed volume ($V$) centered at $\mathbf{X}_0$ is given by

$$\bar{e}_V(\mathbf{\eta}(\mathbf{X}_0), \nabla\mathbf{\eta}) = \frac{1}{V} \sum_i^N U(\mathbf{\eta}(\mathbf{Y}_i), \nabla\mathbf{\eta}), \tag{86}$$

where $U(\mathbf{\eta}(\mathbf{Y}_i), \nabla\mathbf{\eta})$ is potential energy of atom $i$ in configuration $\{Y\}$. Apparently, the value of the average energy density is dependent on the volume because of the presence of strain gradient. It is usually required that the selected volume (characteristic volume) should be able to reflect size effects of interests for a certain problem. In present work, we mainly concern about the higher-order elastic instability of single crystals under dynamic loadings. Thereby, our characteristic volume is the atom volume at certain position (i.e., $\mathbf{X}_0$). Then average energy density over the characteristic volume is

$$\bar{e}_{\Omega_X}(\mathbf{\eta}, \mathbf{\kappa}) = \frac{1}{\Omega_X} U(\mathbf{\eta}(\mathbf{Y}_0), \mathbf{\kappa}) = \frac{1}{\Omega_X}[U_0(\mathbf{\eta}(\mathbf{X}_0), \mathbf{\kappa}) + \Delta U(\mathbf{\eta}, \mathbf{\kappa})], \tag{87}$$

where $\Omega_X$ is the atom volume at $\mathbf{X}_0$, $U_0(\mathbf{\eta}(\mathbf{X}_0), \mathbf{\kappa})$ is potential energy per atom in configuration $\{X\}$. Energy increment ($\Delta U$) after introducing the uniform strain gradient is

$$\Delta U(\mathbf{\eta}, \mathbf{\kappa}) = U(\mathbf{\eta}(\mathbf{Y}_0), \mathbf{\kappa}) - U_0(\mathbf{\eta}(\mathbf{X}_0), \mathbf{\kappa}). \tag{88}$$

During the past few decades, quantities of atomic simulation researches have shown that interatomic interactions of many metals could be well described by embedded-atom method (EAM) potentials. In this work, we will derive the microscopic expressions of the elastic constants for solid binding by modified analytic EAM potential developed by (Wang et al., 2014). The results could be also applied for solids binding by the EAM potential through dropping the energy modified term in the modified analytic EAM potential. According to the modified analytic EAM potential, potential energy at $\mathbf{Y}_0$ is

$$U(\mathbf{\eta}, \mathbf{\kappa}) = \frac{1}{2} \sum_{i \neq 0} \phi(Y_i) + F(\rho_\mathbf{Y}) + M(P_\mathbf{Y}), \tag{89}$$

where

$$\rho_\mathbf{Y} = \sum_{i \neq 0} f(Y_i), \quad P_\mathbf{Y} = \sum_{i \neq 0} g(Y_i). \tag{90}$$

The summations in Eq. (86) and (87) go over all neighbors within a distance of $r_c$ from the central atom. Hereafter, we refer to the distance as cutoff distance. Detailed function form of pairwise interaction $\phi(r)$, atom electron density $f(r)$, spherical deviation term $g(r)$, embedding energy $F(\rho)$ and energy modified term $M(P)$ could be given in either tabulated or analytic form. For the modified analytic EAM potential of iron, analytic function form could be found in (Wang et al., 2014). However, the tabulated forms are more transferable among different atomic simulation software and thus widely adopted by the EAM potentials. Below, our derivations are irrespective with the detailed function form of EAM potentials.

To evaluate the energy increment due to the uniform strain gradient, we expand $U(\boldsymbol{\eta}(\mathbf{Y}_0), \boldsymbol{\kappa})$ at configuration $\{\mathbf{X}\}$ to the second order in terms of displacements of atoms within the characteristic volume, that is,

$$\Delta U(\boldsymbol{\eta}, \boldsymbol{\kappa}) = \frac{1}{2}\sum_{i \neq 0}[\phi(\mathbf{Y}_i) - \phi(\mathbf{X}_i)] + F(\rho_0(\{\mathbf{Y}_m\})) - F(\rho_0(\{\mathbf{X}_m\}))$$

$$+ M(P_0(\{\mathbf{Y}_m\})) - M(P_0(\{\mathbf{X}_m\}))$$

$$= \frac{1}{2}\sum_{i \neq 0} \Delta\phi(\mathbf{Y}_i) + \Delta F(\{\mathbf{Y}_m\}) + \Delta M(\{\mathbf{Y}_m\}), \qquad (91)$$

where

$$\Delta\phi(\mathbf{Y}_i) = \phi(\mathbf{Y}_i) - \phi(\mathbf{X}_i)$$

$$= \phi'(X_i)\frac{X_i^\alpha}{X_i}u_i^\alpha + \frac{1}{2}\left[\phi''(X_i)\frac{X_i^\alpha X_i^\beta}{X_i^2} + \phi'(X_i)\left(\frac{1}{X_i}\delta_{\alpha\beta} - \frac{X_i^\alpha X_i^\beta}{X_i^3}\right)\right]u_i^\alpha u_i^\beta, \qquad (92)$$

$$\Delta F(\{\mathbf{Y}_m\}) = \frac{\partial F}{\partial \rho_0}\Delta\rho_0 + \frac{1}{2}\frac{\partial^2 F}{\partial \rho_0^2}(\Delta\rho_0)^2 \qquad (93)$$

$$\Delta M(\{\mathbf{Y}_m\}) = \frac{\partial P}{\partial g_0}\Delta g_0 + \frac{1}{2}\frac{\partial^2 P}{\partial g_0^2}(\Delta g_0)^2 \qquad (94)$$

In above expansions, we have assumed that the displacement of central atom is zero, i.e., $\mathbf{u}_0 = \mathbf{Y}_0 - \mathbf{X}_0 = 0$. This assumption could be always satisfied through adjusting the strain of configuration $\{\mathbf{X}\}$ to the exact value of $\boldsymbol{\eta}(\mathbf{Y}_0)$ and taking the adjusted configuration as the reference one. Moreover, the cutoff distance is small quantities (the same order as lattice constant). Thus, the displacements of all neighbors within the cutoff distance, i.e., $\mathbf{u}_i$, are small quantities, which forms the bases of the expansions in Eq. (91)-(94). Similarly, increments of $\rho_0$ and $g_0$ due to the strain gradient are expanded to the second order of displacements, that is

$$\Delta\rho_0 = \sum_{i \neq 0}[f(\mathbf{Y}_i) - f(\mathbf{X}_i)] = \sum_{i \neq 0}\left[\frac{\partial f}{\partial x_i^\alpha}u_i^\alpha + \frac{1}{2}\frac{\partial^2 f}{\partial x_i^\alpha \partial x_i^\beta}u_i^\alpha u_i^\beta\right]$$

$$= \sum_{i \neq 0}\left[f'(X_i)\frac{X_i^\alpha}{X_i}u_i^\alpha + \frac{1}{2}\left(f''(X_i)\frac{X_i^\alpha X_i^\beta}{X_i^2} + f'(X_i)\left(\frac{1}{X_i}\delta_{\alpha\beta} - \frac{X_i^\alpha X_i^\beta}{X_i^3}\right)\right)u_i^\alpha u_i^\beta\right], \qquad (95)$$

$$\Delta g_0 = \sum_{i \neq 0}[g(\mathbf{Y}_i) - g(\mathbf{X}_i)] = \sum_{i \neq 0}\left[\frac{\partial g}{\partial x_i^\alpha}u_i^\alpha + \frac{1}{2}\frac{\partial^2 g}{\partial x_i^\alpha \partial x_i^\beta}u_i^\alpha u_i^\beta\right]$$

$$= \sum_{i \neq 0}\left[g'(X_i)\frac{X_i^\alpha}{X_i}u_i^\alpha + \frac{1}{2}\left(g''(X_i)\frac{X_i^\alpha X_i^\beta}{X_i^2} + g'(X_i)\left(\frac{1}{X_i}\delta_{\alpha\beta} - \frac{X_i^\alpha X_i^\beta}{X_i^3}\right)\right)u_i^\alpha u_i^\beta\right]. \qquad (96)$$

Substituting Eq. (95) and (96) into (93) and (94), respectively, we get

$$\Delta F(\{\mathbf{Y}_m\}) = F'(\rho_0) \sum_{i\neq 0} \left[ f'(X_i) \frac{X_i^\alpha}{X_i} u_i^\alpha + \frac{1}{2}\left( f''(X_i) \frac{X_i^\alpha X_i^\beta}{X_i^2} + f'(X_i) \left( \frac{1}{X_i} \delta_{\alpha\beta} - \frac{X_i^\alpha X_i^\beta}{X_i^3} \right) \right) u_i^\alpha u_i^\beta \right]$$

$$+ \frac{1}{2} F''(\rho_0) \sum_{i\neq 0} \sum_{j\neq 0} f'(X_i) f'(X_j) \frac{X_i^\alpha X_j^\beta}{X_i X_j} u_i^\alpha u_j^\beta, \tag{97}$$

$$\Delta M(\{\mathbf{Y}_m\}) = \frac{1}{N_0} \sum_i [M(P_i(\{Y\})) - M(P_i(\{X\}))]$$

$$= M'(P_0) \sum_{i\neq 0} \left[ g'(X_i) \frac{X_i^\alpha}{X_i} u_i^\alpha + \frac{1}{2}\left( g''(X_i) \frac{X_i^\alpha X_i^\beta}{X_i^2} + g'(X_i) \left( \frac{1}{X_i} \delta_{\alpha\beta} - \frac{X_i^\alpha X_i^\beta}{X_i^3} \right) \right) u_i^\alpha u_i^\beta \right]$$

$$+ \frac{1}{2} M''(P_0) \sum_{i\neq 0} \sum_{j\neq 0} g'(X_i) g'(X_j) \frac{X_i^\alpha X_j^\beta}{X_i X_j} u_i^\alpha u_j^\beta. \tag{98}$$

With Eq. (92), (97) and (98), the energy increment could be rewritten as

$$\Delta U(\boldsymbol{\eta}, \boldsymbol{\kappa}) = \frac{1}{2} \sum_{i\neq 0} \left\{ \phi'(X_i) \frac{X_i^\alpha}{X_i} u_i^\alpha + \frac{1}{2}\left[ \phi''(X_i) \frac{X_i^\alpha X_i^\beta}{X_i^2} + \phi'(X_i) \left( \frac{1}{X_i} \delta_{\alpha\beta} - \frac{X_i^\alpha X_i^\beta}{X_i^3} \right) \right] u_i^\alpha u_i^\beta \right\}$$

$$+ \sum_{i\neq 0} F'(\rho_0) f'(X_i) \frac{X_i^\alpha}{X_i} u_i^\alpha + \frac{1}{2} \sum_{i\neq 0} \sum_{j\neq 0} F''(\rho_0) f'(X_i) f'(X_j) \frac{X_i^\alpha X_j^\beta}{X_i X_j} u_i^\alpha u_j^\beta$$

$$+ \frac{1}{2} \sum_{i\neq 0} F'(\rho_0) \left( f''(X_i) \frac{X_i^\alpha X_i^\beta}{X_i^2} + f'(X_i) \left( \frac{1}{X_i} \delta_{\alpha\beta} - \frac{X_i^\alpha X_i^\beta}{X_i^3} \right) \right) u_i^\alpha u_i^\beta$$

$$+ \sum_{i\neq 0} M'(P_0) g'(X_i) \frac{X_i^\alpha}{X_i} u_i^\alpha + \frac{1}{2} \sum_{i\neq 0} \sum_{j\neq 0} M''(P_0) g'(X_i) g'(X_j) \frac{X_i^\alpha X_j^\beta}{X_i X_j} u_i^\alpha u_j^\beta$$

$$+ \frac{1}{2} \sum_{i\neq 0} M'(P_0) \left( g''(X_i) \frac{X_i^\alpha X_i^\beta}{X_i^2} + g'(X_i) \left( \frac{1}{X_i} \delta_{\alpha\beta} - \frac{X_i^\alpha X_i^\beta}{X_i^3} \right) \right) u_i^\alpha u_i^\beta$$

$$= \sum_{i\neq 0} [\phi'(X_i)/2 + F'f'(X_i) + M'g'(X_i)] \frac{X_i^\alpha}{X_i} u_i^\alpha$$

$$+ \frac{1}{2} \sum_{i\neq 0} \left\{ \frac{1}{2}\left[ \phi''(X_i) \frac{X_i^\alpha X_i^\beta}{X_i^2} + \phi'(X_i) \left( \frac{1}{X_i} \delta_{\alpha\beta} - \frac{X_i^\alpha X_i^\beta}{X_i^3} \right) \right] + F' \left[ f''(X_i) \frac{X_i^\alpha X_i^\beta}{X_i^2} \right. \right.$$

$$\left. + f'(X_i) \left( \frac{1}{X_i} \delta_{\alpha\beta} - \frac{X_i^\alpha X_i^\beta}{X_i^3} \right) \right]$$

$$\left. + M' \left[ g''(X_i) \frac{X_i^\alpha X_i^\beta}{X_i^2} + g'(X_i) \left( \frac{1}{X_i} \delta_{\alpha\beta} - \frac{X_i^\alpha X_i^\beta}{X_i^3} \right) \right] \right\} u_i^\alpha u_i^\beta$$

$$+ \frac{1}{2} \sum_{i\neq 0} \sum_{j\neq 0} [F''f'(X_i)f'(X_j) + M''g'(X_i)g'(X_j)] \frac{X_j^\beta}{X_j} \frac{X_i^\alpha}{X_i} u_i^\alpha u_j^\beta$$

$$\equiv \sum_{i\neq 0} P_i^\alpha u_i^\alpha + \frac{1}{2} \sum_{i\neq 0} V_i^{\alpha\beta} u_i^\alpha u_i^\beta + \frac{1}{2} \sum_{i\neq 0} \sum_{j\neq 0} H_{ij}^{\alpha\beta} u_i^\alpha u_j^\beta, \tag{99}$$

where

$$P_i^\alpha = [\phi'(X_i)/2 + F'f'(X_i) + M'g'(X_i)] \frac{X_i^\alpha}{X_i}, \tag{100}$$

$$V_i^{\alpha\beta} = \frac{1}{2}\left[\phi''(X_i)\frac{X_i^\alpha X_i^\beta}{X_i^2} + \phi'(X_i)\left(\frac{1}{X_i}\delta_{\alpha\beta} - \frac{X_i^\alpha X_i^\beta}{X_i^3}\right)\right]$$

$$+ F'\left[f''(X_i)\frac{X_i^\alpha X_i^\beta}{X_i^2} + f'(X_i)\left(\frac{1}{X_i}\delta_{\alpha\beta} - \frac{X_i^\alpha X_i^\beta}{X_i^3}\right)\right]$$

$$+ M'\left[g''(X_i)\frac{X_i^\alpha X_i^\beta}{X_i^2} + g'(X_i)\left(\frac{1}{X_i}\delta_{\alpha\beta} - \frac{X_i^\alpha X_i^\beta}{X_i^3}\right)\right], \tag{101}$$

$$H_{ij}^{\alpha\beta} = \left[F''f'(X_i)f'(X_j) + M''g'(X_i)g'(X_j)\right]\frac{X_i^\alpha}{X_i}\frac{X_j^\beta}{X_j}. \tag{102}$$

Substituting Eq. (83) into (102) and making use of $u_\alpha(0) = 0$, we have

$$\Delta U(\boldsymbol{\eta}, \boldsymbol{\kappa}) = \sum_{i \neq 0} P_i^\alpha \left(u_{\alpha,\mu} X_i^\mu + \frac{1}{2} u_{\alpha,\mu\nu} X_i^\mu X_i^\nu\right)$$

$$+ \frac{1}{2}\sum_{i \neq 0} V_i^{\alpha\beta} \left(u_{\alpha,\mu} X_i^\mu + \frac{1}{2} u_{\alpha,\mu\nu} X_i^\mu X_i^\nu\right)\left(u_{\beta,\lambda} X_i^\lambda + \frac{1}{2} u_{\beta,\lambda\rho} X_i^\lambda X_i^\rho\right)$$

$$+ \frac{1}{2}\sum_i \sum_j H_{ij}^{\alpha\beta} \left(u_{\alpha,\mu} X_i^\mu + \frac{1}{2} u_{\alpha,\mu\nu} X_i^\mu X_i^\nu\right)\left(u_{\beta,\lambda} X_j^\lambda + \frac{1}{2} u_{\beta,\lambda\rho} X_j^\lambda X_j^\rho\right)$$

$$= \left(\sum_{i \neq 0} P_i^\alpha X_i^\mu\right) u_{\alpha,\mu} + \frac{1}{2}\left(\sum_{i \neq 0} P_i^\alpha X_i^\mu X_i^\nu\right) u_{\alpha,\mu\nu}$$

$$+ \frac{1}{2}\left(\sum_{i \neq 0} V_i^{\alpha\beta} X_i^\mu X_i^\lambda + \sum_i \sum_j H_{ij}^{\alpha\beta} X_i^\mu X_j^\lambda\right) u_{\alpha,\mu} u_{\beta,\lambda}$$

$$+ \frac{1}{4}\left(\sum_{i \neq 0} V_i^{\alpha\beta} X_i^\mu X_i^\lambda X_i^\rho + \sum_i \sum_j H_{ij}^{\alpha\beta} X_i^\mu X_j^\lambda X_j^\rho\right) u_{\alpha,\mu} u_{\beta,\lambda\rho}$$

$$+ \frac{1}{4}\left(\sum_{i \neq 0} V_i^{\alpha\beta} X_i^\mu X_i^\nu X_i^\lambda + \sum_i \sum_j H_{ij}^{\alpha\beta} X_i^\mu X_i^\nu X_j^\lambda\right) u_{\alpha,\mu\nu} u_{\beta,\lambda}$$

$$+ \frac{1}{8}\left(\sum_{i \neq 0} V_i^{\alpha\beta} X_i^\mu X_i^\nu X_i^\lambda X_i^\rho + \sum_i \sum_j H_{ij}^{\alpha\beta} X_i^\mu X_i^\nu X_j^\lambda X_j^\rho\right) u_{\alpha,\mu\nu} u_{\beta,\lambda\rho}$$

$$= \tilde{P}_{\alpha\mu} u_{\alpha,\mu} + \tilde{Q}_{\alpha\mu\nu} u_{\alpha,\mu\nu} + \frac{1}{2}\tilde{C}_{\alpha\mu\beta\lambda} u_{\alpha,\mu} u_{\beta,\lambda} + \tilde{V}_{\alpha\mu\beta\lambda\rho} u_{\alpha,\mu} u_{\beta,\lambda\rho}$$

$$+ \frac{1}{2}\tilde{H}_{\alpha\mu\nu\beta\lambda\rho} u_{\alpha,\mu\nu} u_{\beta,\lambda\rho}, \tag{103}$$

where

$$\tilde{P}_{\alpha\mu} = \sum_{i \neq 0} P_i^\alpha X_i^\mu, \tag{104}$$

$$\tilde{Q}_{\alpha\mu\nu} = \frac{1}{2}\sum_{i \neq 0} P_i^\alpha X_i^\mu X_i^\nu, \tag{105}$$

$$\tilde{C}_{\alpha\mu\beta\lambda} = \sum_{i \neq 0} V_i^{\alpha\beta} X_i^\mu X_i^\lambda + \sum_i \sum_j H_{ij}^{\alpha\beta} X_i^\mu X_j^\lambda, \tag{106}$$

$$\tilde{V}_{\alpha\mu\beta\lambda\rho} = \frac{1}{4}\left(2\sum_{i \neq 0} V_i^{\alpha\beta} X_i^\mu X_i^\lambda X_i^\rho + \sum_i \sum_j H_{ij}^{\alpha\beta} X_i^\mu X_j^\lambda X_j^\rho + \sum_i \sum_j H_{ij}^{\alpha\beta} X_i^\lambda X_i^\rho X_j^\mu\right), \tag{107}$$

$$\tilde{H}_{\alpha\mu\nu\beta\lambda\rho} = \frac{1}{4}\left(\sum_{i \neq 0} V_i^{\alpha\beta} X_i^\mu X_i^\nu X_i^\lambda X_i^\rho + \sum_i \sum_j H_{ij}^{\alpha\beta} X_i^\mu X_i^\nu X_j^\lambda X_j^\rho\right). \tag{108}$$

Using Eq. (84) and (85) to express Eq. (103) in terms of stain and strain gradients and omitting terms higher than the second order, we get

$$\Delta U(\boldsymbol{\eta}, \boldsymbol{\kappa}) = \Omega_{\mathbf{X}} \left( \sigma_{\alpha\mu}\eta_{\alpha\mu} + \frac{1}{2}C_{\alpha\mu\beta\lambda}\eta_{\alpha\mu}\eta_{\beta\lambda} + \tau_{\alpha\mu\nu}\eta_{\alpha\mu,\nu} + W_{\alpha\mu\beta\lambda\rho}\eta_{\alpha\mu}\eta_{\beta\lambda,\rho} \right.$$
$$\left. + \frac{1}{2}\Pi_{\alpha\mu\upsilon\beta\lambda\rho}\eta_{\alpha\mu,\nu}\eta_{\beta\lambda,\rho} \right)$$

(109)

where

$$\sigma_{\alpha\mu} = \tilde{P}_{\alpha\mu}/\Omega_{\mathbf{X}}, \tag{110}$$
$$C_{\alpha\mu\beta\lambda} + C_{\alpha\mu\lambda\beta} + C_{\mu\alpha\beta\lambda} + C_{\mu\alpha\lambda\beta} = 4\left(\tilde{C}_{\alpha\mu\beta\lambda} - \delta_{\alpha\beta}\tilde{P}_{\mu\lambda}\right)/\Omega_{\mathbf{X}}, \tag{111}$$
$$\tau_{\alpha\mu\nu} = \tilde{Q}_{\alpha\mu\nu}/\Omega_{\mathbf{X}}, \tag{112}$$
$$W_{\alpha\mu\beta\lambda\rho} + W_{\mu\alpha\beta\lambda\rho} + W_{\alpha\mu\lambda\beta\rho} + W_{\mu\alpha\lambda\beta\rho} = 4\left(\tilde{V}_{\alpha\mu\beta\lambda\rho} - \tilde{Q}_{\lambda\mu\rho}\delta_{\alpha\beta}\right)/\Omega_{\mathbf{X}}, \tag{113}$$
$$\Pi_{\alpha\mu\upsilon\beta\lambda\rho} + \Pi_{\alpha\mu\upsilon\lambda\beta\rho} + \Pi_{\mu\alpha\upsilon\beta\lambda\rho} + \Pi_{\mu\alpha\upsilon\lambda\beta\rho} = 4\tilde{H}_{\alpha\mu\upsilon\beta\lambda\rho}/\Omega_{\mathbf{X}}. \tag{114}$$

Due to the symmetries of ($\alpha\leftrightarrow\mu$), ($\beta\leftrightarrow\lambda$) and ($\alpha\mu$)↔($\beta\lambda$), (111), (113) and (114) could be rewritten as

$$\Omega_{\mathbf{X}}C_{\alpha\mu\beta\lambda} = \tilde{C}_{\alpha\mu\beta\lambda} - \delta_{\alpha\beta}\tilde{P}_{\mu\lambda} = \tilde{C}_{\alpha\mu\beta\lambda} - \Omega_{\mathbf{X}}\delta_{\alpha\beta}\sigma_{\mu\lambda}, \tag{111'}$$
$$\Omega_{\mathbf{X}}W_{\alpha\mu\beta\lambda\rho} = \tilde{V}_{\alpha\mu\beta\lambda\rho} - \tilde{Q}_{\lambda\mu\rho}\delta_{\alpha\beta} = \tilde{V}_{\alpha\mu\beta\lambda\rho} - \Omega_{\mathbf{X}}\delta_{\alpha\beta}\tau_{\lambda\mu\rho}, \tag{113'}$$
$$\Omega_{\mathbf{X}}\Pi_{\alpha\mu\upsilon\beta\lambda\rho} = \tilde{H}_{\alpha\mu\upsilon\beta\lambda\rho}. \tag{114'}$$

Notably, **C** and **W** depend on stress (**σ**) and higher order stress (**τ**), respectively. This is because the reference configuration, i.e. {**X**}, does not necessarily stays at equilibrium state where the stress and the higher order stress are zeros. If the potential energy density is expanded in terms of small linear strain and the corresponding strain gradient, **C** and **W** would not depend on the stress and higher order stress, that is

$$\Omega_{\mathbf{X}}C_{\alpha\mu\beta\lambda} = \tilde{C}_{\alpha\mu\beta\lambda}, \tag{111''}$$
$$\Omega_{\mathbf{X}}W_{\alpha\mu\beta\lambda\rho} = \tilde{V}_{\alpha\mu\beta\lambda\rho}. \tag{113''}$$

Through substituting Eq. (109) into (87), potential energy density of configuration {**Y**} at $\mathbf{Y}_0$ is

$$e_{\Omega_X}(\boldsymbol{\eta}(\mathbf{Y}_0), \boldsymbol{\kappa}) = \frac{1}{\Omega_X}[U_0 + \Delta U(\boldsymbol{\eta}, \boldsymbol{\kappa})]$$

$$= e_0 + \sigma_{\alpha\mu}\eta_{\alpha\mu} + \frac{1}{2}C_{\alpha\mu\beta\lambda}\eta_{\alpha\mu}\eta_{\beta\lambda} + \tau_{\alpha\mu\nu}\eta_{\alpha\mu,\nu} + W_{\alpha\mu\beta\lambda\rho}\eta_{\alpha\mu}\eta_{\beta\lambda,\rho} + \frac{1}{2}\Pi_{\alpha\mu\upsilon\beta\lambda\rho}\eta_{\alpha\mu,\nu}\eta_{\beta\lambda,\rho},$$

(115)

where $e_0$ is energy density of the reference configuration. Thereby, microscopic expressions of the elastic constants, defined by Eq. (25)-(29), are given by (110)-(114), respectively. Specially, from the expression for $\Pi_{\alpha\mu\upsilon\beta\lambda\rho}$ (See Eq. 114', 108, 102 and 101), we find that $\rho$ and $\upsilon$ (or $\alpha$ and $\beta$, $\mu$ and $\lambda$) are exchangeable due to spherically average approximation to electron density adopted by the EAM potential models. Such symmetries for Π may be lost for solids binding through angular-dependence interatomic interaction potentials. Anyhow, the symmetry of $\rho\leftrightarrow\upsilon$ (or $\alpha\mu\leftrightarrow\beta\lambda$), leading to the relationship of (74), gives us a good starting point for understanding the interactions between mechanical instabilities triggered by strain and strain gradient. Additionally, it should be noticed that our derivations do not assume that the solid should be a crystal at 0 K. That is to say, in principle, the microscopic expressions of the elastic constants could be applied for any solids at all temperature below their melting points. However, the calculated elastic constants are "local" (position-dependence) and rely on the characteristic size (volume). For example, when a crystal is in thermodynamic equilibrium at nonzero temperature, the calculated elastic constants will depend on the position of the characteristic volume due to displacement fluctuations of atoms. In this case, a characteristic volume, over which the elastic constants are averaged, should be chosen so that the elastic constants are not sensitive to the position, i.e., the fluctuations of atom displacements. The expressions could also be used to calculate the elastic constants for amorphous materials through choosing a proper characteristic volume. Specially, when the target system is in non-equilibrium state, for example, solids under dynamic loadings, the calculated elastic constants represent local quantities characterizing the mechanical properties of local materials. In the next part, we will use the local elastic constants to study local mechanical instabilities of metals under dynamic loadings. In fact, the elastic constants calculated by Eq. (110)-(114) are corresponding to ones at current configuration with or without initial deformation. Thus, these expressions could be employed to study mechanical instabilities of a solid at arbitrary finite strain states through taking the strained solid as current configuration and using stability criteria (36) developed for small strains.

# 6.  Mechanical Instabilities of Metals under Dynamic Loadings

## 6.1 Mechanical Instabilities of Copper and Aluminum Single Crystals under Ramp Compressions

Strain gradient induced mechanical instabilities are recently reported in iron single crystals under ramp compressions simulated by NEMD simulations (Wang et al., 2017). It is found that singularities would arise in wave profile when the instabilities take place in the loaded iron samples. In present work, the ramp compression technique continues to be adopted for studying mechanical instabilities of two typical plastic metals (copper and aluminum). Interatomic interactions of copper and aluminum are described by embedded atom model potential of (Mishin et al., 1999; Mishin et al., 2001), which is suitable for high pressure applications. A copper single crystals, with initial sizes of 18.08×18.08×289.20 nm, are impacted along +Z direction (corresponding to [001] direction) through a moving infinite massive piston at 0K. Ramp compression is generated via linearly increasing the impacting velocity ($v_p$) of the piston from zero to a maximum value ($v_p^{max}$) within a given time ($t_{rising}$). After the ramp compression, the piston keeps its maximum velocity for a certain time ($t_s$) before being removed away from compressed sample. Applied strain rate could be evaluated by $v_p^{max}/(c_L t_{rising})$, where $c_L$ is longitudinal sonic speed along compression direction. The values of $c_L$ for metals involved in present work are listed in Table I. Our simulated strain rates range from $10^9$ to $10^{10}$ s$^{-1}$. More detailed settings could be found in *Supplementary materials*. For aluminum, a single crystal sample, with an initial size of 20.15×20.15×324.00 nm, are employed for the ramp compressions, with $v_{max}$ = 2.0 km/s and $t_{rising}$ = 80 ps, along [001] direction. Besides, lattice deformations of compressed samples are analyzed by the lattice analyses technique mentioned in ref. (Wang et al., 2015). If not specified, finite Lagrangian strain (**e**) and the corresponding strain gradient are employed in the deformation analyses of present work. In the remainder of this work, the summation convention over repeated indexes is canceled.

According to Eq. (50), applied strain gradient ($e_{33,3}$) of the ramp compressions could be approximately estimated, to the second order, by

$$e_{33,3} \approx \varepsilon_{33,3} + \frac{1}{2}\sum_k \left(\varepsilon_{k3,3}\varepsilon_{k3} + \varepsilon_{k3}\varepsilon_{k3,3}\right) = \varepsilon_{33,3} + \varepsilon_{33}\varepsilon_{33,3}, \tag{116}$$

where only uniaxial strain along Z direction is considered. It should be noted that the **η** and its gradient in the Eq. (50) are zeros since small linear strain (ε), defined here, is measured from configuration $\{a\}$. And gradient of the linear strain is

$$\varepsilon_{33,3} = \frac{d\varepsilon_{33}}{dZ} = \frac{d}{dZ}\frac{du_Z}{dZ} = \frac{dv}{c_L dZ} = \frac{a_Z}{c_L^2}, \tag{117}$$

where $a_Z = v_p^{max}/t_{rising}$ is average acceleration of particles. Taking the ramp compression on Cu with $v_p^{max} = 2km/s$ and $t_{rising} = 80ps$, the ramp on Al with $v_p^{max} = 2km/s$ and $t_{rising} = 80ps$ and the ramp on Fe with $v_p^{max} = 0.8km/s$ and $t_{rising} = 15ps$ for example, the ranges of the applied strain gradient for the three cases are estimated to be [0.20, 0.32], [0.073, 0.097] and [0.24, 0.28], respectively, where units are $1\times10^{-3}$ Å$^{-1}$.

Wave profiles represented by particle velocity and strain for copper are shown in Fig. 1. The results in Fig. 1 indicate that the copper sample could be elastically compressed to a strain much higher than a critical strain (marked by "B" in the figure), predicted by modified Born criteria, before plasticity takes place. This over-pressurization is also observed in iron, which could be well interpreted by nucleation time of the plasticity or phase transition arising from intrinsic vibrational period of lattice atoms. The nucleation of plasticity or phase transition involves atom rearrangements which are a result of quantities of basic lattice events, i.e., atom jump from one position to its neighboring. And each lattice event is triggered by the one among hundreds or thousands of the atom vibrations around their equilibrium positions. Thereby, strain rate effect arises when the strain rate is comparable to phonon frequency of a crystal. Typical phonon frequency of metals is about 1THz, corresponding to a strain rate of about $10^{12}$ s$^{-1}$, which is very approachable to the simulated strain rates and thus, explains the over-pressurization. Besides, average number of the atom vibrations needed to generate a lattice event is adequately smaller in an instable solid than that in a stable solid. Namely, inelasticity takes place more rapidly in the instable solid than a stable one. Consequently, in the wave profiles, singularities firstly arise at the instable region during propagations of ramp waves. In Fig. 1, positions of the singularities are marked by I and II, where strains are about -0.042 and -0.069, respectively. The strains are the critical strains at which instabilities takes place. For ramp compressions with the other applied strain rates investigated in present work, the critical strains are very close to the ones given above. This would be clarified later in Part 6.2.

As a complementation, mechanical instabilities of aluminum single crystal is also investigated using the ramp compression technique. The analyses for copper could also be applied for aluminum. However, as shown Fig. 3, three additional instabilities, marked by I, II and III, would take place before strain instability (marked by "B") occurs. The critical strains for I, II and III are -0.013, -0.042 and -0.089. Besides, III are not well defined since knee at III is not as obvious as that at I and II. As shown in Fig. 3b, the critical strain at III lies within a range from -0.065 to -0.1. Additionally, the critical strain of strain gradient instability for iron is observed to be about 0.09 (Wang et al., 2017). The critical strains for the three metals will be consistently explained in the next part by the instability condition developed in Part 3 and 4.

## 6.2 Strain-Gradient Related Elastic Constants and Mechanical Instabilities of Copper, Aluminum and Iron

Alternatively, the critical strains could be predicted directly using the stability conditions developed in Part 2. For uniaxial strain and strain gradient along Z direction, we have

$$\hat{\Theta}_{IJKL} = \frac{1}{2}\hat{\Pi}_{IJ3KL3,33} = \frac{1}{2}\frac{\partial^2 \hat{\Pi}_{IJ3KL3}}{\partial \eta_{33}^2}\eta_{33,3}^2. \tag{118}$$

And the Birch coefficients ($\hat{\mathbf{B}}$) at arbitrary strains could be calculated by (65), or more symmetrically (Wang et al., 1995; Wang et al., 1993), by

$$\hat{B}_{IJKL} = \hat{C}_{IJKL} + \frac{1}{2}\left(\hat{\sigma}_{IL}\delta_{JK} + \hat{\sigma}_{JL}\delta_{IK} + \hat{\sigma}_{IK}\delta_{JL} + \hat{\sigma}_{JK}\delta_{IL} - 2\hat{\sigma}_{IJ}\delta_{KL}\right), \tag{119}$$

where the elastic constants and stresses are calculated by the Eq. (111'') statically. According to discussions in Part 3 and Part 4, mechanical instability due to disturbances of strain and strain gradient begins when the minimum eigenvalue ($\hat{K}_{min}$) of $\hat{\mathbf{K}} = \hat{\mathbf{B}} + \frac{1}{2}\hat{\mathbf{\Theta}}$ becomes negative. With the Eq. (114'), (119) and (111''), $\hat{\Pi}$, as well as $\hat{\mathbf{B}}$, for solids at arbitrary strains could be calculated using a molecular statics code. For convenience, indexes $IJ$ (and $KL$) in $\hat{\Pi}_{IJMKLN}$ are contracted into one using Voigt convention. As shown in Table II, typical magnitude of $\hat{\Pi}_{IJMKLN}$ is about $1\times10^2$ GPa·Å$^2$, which is comparable to the one given by (Maranganti and Sharma, 2007). Further, according to Eq. (118) and (119), $\hat{K}_{min}$ is calculated at different uniaxial strain and strain gradient for copper, aluminum and iron. The calculated results are shown in Fig. 3 and Fig. 4. For copper, three instable regions, separated by gray contour lines, are marked by I, II' and II'' in $\varepsilon_{33}$-$\varepsilon_{33,3}$ diagram. The first two instable regions arise from strain gradient instabilities and region II'' is mainly caused by strain instabilities. Region II' are very close to the II'' at small strain gradients and will intersect with II'' at large strain gradients. The critical strains at I and II in Fig. 1, i.e., -0.042 and -0.069, respectively, are well located at lower boundaries of the instable region I and II' (See Fig. 3). However, minimum strain gradient required to trigger the strain gradient instability is about 0.0074 which is larger than the applied strain gradient estimated in Part 6.1. Similar phenomena are also observed in aluminum and iron. This is because actual strain gradient in compressed samples is increasing with the steepening of wave profiles during wave propagations although its initial value is equal to the applied strain gradient. On one hand, the actual strain gradient is determined by the degree of the steepening which arises from nonlinear elasticity at large strains. On the other hand, the steepening in elastic waves would end when strain instabilities begin. Thereby, when the initial applied strain gradient is not much smaller than the minimum strain gradient, the actual strain gradient may also increase to a value larger than the minimum strain gradient and trigger strain gradient instabilities before the strain instabilities may take place.

Using the $\varepsilon_{33}$-$\varepsilon_{33,3}$ diagram in Fig. 4, similar analyses could also be employed to explain the critical strains observed in aluminum and iron. Specially, some different characteristics are observed in the $\varepsilon_{33}$-$\varepsilon_{33,3}$ diagram of aluminum. Firstly, critical strain for the instability region I'' is not observed from the wave profiles in Fig. 2. This is because region I'' is so close to I' that the knee in the wave profiles for the two regions are distinguishable. As shown in Fig. 4a, the difference in the critical strain between the two regions is about 0.008 which will generate a separation between the corresponding knees by 0.8 nm under an applied strain gradient of $10^{-3}$ Å$^{-1}$. This separation distance is too small to be distinguished in the wave profiles shown in Fig. 2. Secondly, strain gradient instabilities take place at region I', I'', II and III', while strain instabilities happen at region III''. And the region III' consists of a wide band with a smeared boundary, while the region I, I' and II consist of narrow "sharp" bands. The smeared boundary means that the value of $\hat{K}_{min}$ nearby the boundary varies slowly with strain. Since mechanical instabilities grow with the decreasing of $\widehat{K}_{min}$, the smeared boundary will lead to uncertainties for judging the onset of the instabilities through the ramp compression technique. Thereby, the observed critical strain for III in Fig. 2 is hard to be precisely determined. Albeit with the uncertainties of the critical strain, its variation range is almost the same as the strain range covered by region III'. From the discussion above, it could be concluded that the more negative $\widehat{K}_{min}$ is at the instability boundary, the larger a curvature is at the knee of elastic waves.

## 7. Conclusion and Remark

Traditional theories on mechanical instabilities (Elliott et al., 2006a; Elliott et al., 2006b; Wang et al., 1995; Wang et al., 1993) does not consider contributions from strain gradient. However, the strain gradient effects are found to be important in nanomaterials and solids under dynamic loadings. In this work, a strain-gradient related higher order elastic theory is established for the mechanical instabilities through expanding free energy density into a quadratic function of both strain and strain gradient. The theoretical framework is proved to be compatible with the original strain gradient elastic theory proposed by (Mindlin, 1965) and (DiVincenzo, 1986). Because strain gradients are often accompanied by finite strains in many cases, equilibrium equation and stability condition are established for both small and finite strains. By taking solids at small strains as a special case of the ones at finite strains, linear higher order elastic constitutive relationships are consistently obtained. Since stabilities of solids at finite strains could be alternatively interpreted by stabilities of the solids at small strains in current deformation configuration, the latter one could reproduce the same stability condition as the former only if the first order mixed-term vanishes. Previously, this result can only be obtained in centrosymmetric materials. In fact, the result is valid for any solids under the second order approximations to energy in terms of strain and strain gradient. To justify the established theory, we redevelop the theory from atom level and obtain microscopic expressions for the related elastic constants. Unlike previous dynamic approaches proposed by (DiVincenzo, 1986; Maranganti and Sharma, 2007), the elastic constants could be directly calculated via a molecular statics procedure. Specially, the higher order stress could also be calculated by the microscopic expressions, which is still not determined yet at atom level. Finally, mechanical instabilities of three metals, i.e., copper, aluminum and iron, are investigated using the stability conditions where the related elastic constants are directly determined at atom level. The predicted critical strains at onset of the mechanical instabilities agree well with results from NEMD simulations. Thereby, to some extent, at least for crystals, the established higher order theory is equivalent to the widely used empirical-potential-based atomic simulation method.

## Acknowledgements

This work is supported by the National Natural Science Foundation of China (NSFC-NSAF 11076012) and China Postdoctoral Science Foundation (No. 2017M610824).

Table I. Longitudinal sonic speeds of several single crystals along [001] direction

| Materials | Cu | Al | Fe |
|---|---|---|---|
| $c_L$ (km/s) | 3.50 | 5.87 | 4.69 |

Table II. Independent strain-gradient related elastic constants for three metals at zero strain and temperature.

| | $\widehat{\Pi}_{1111}$ | $\widehat{\Pi}_{2121}$ | $\widehat{\Pi}_{4141}$ | $\widehat{\Pi}_{5151}$ | $\widehat{\Pi}_{1121}$ | $\widehat{\Pi}_{2131}$ | $\widehat{\Pi}_{4152}$ | $\widehat{\Pi}_{1162}$ |
|---|---|---|---|---|---|---|---|---|
| Cu (GPa·Å$^2$) | 0.0 | 160.7 | 0.0 | 173.5 | 151.2 | 0.0 | 0.0 | 155.9 |
| Al (GPa·Å$^2$) | 210.2 | 0.0 | 0.0 | 0.0 | 0.0 | 0.0 | 0.0 | 0.0 |
| Fe (GPa·Å$^2$) | 11174.6 | 123.2 | 178.4 | 134.8 | 125.8 | 179.7 | 179.0 | 124.5 |

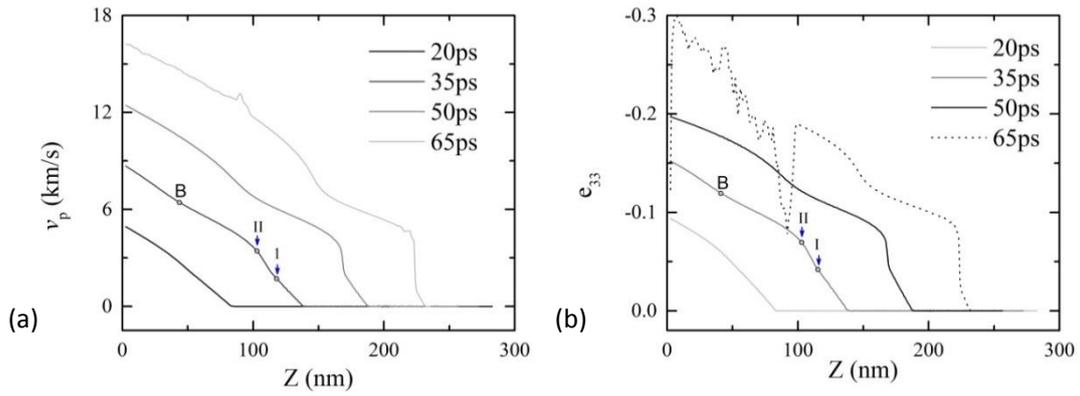

Fig. 1. Wave profile of (a) particle velocity and (b) strain for copper single crystal under ramp compressions, with $v_{max}$ = 2.0 km/s and $t_{rising}$ = 80 ps, along [001] direction. Plasticity takes place in the sample at about 65ps, before which only elastic compressions are observed. At 35ps, initial knees (marked by I and II) in the elastic waves are created by mechanical instabilities. And the knee would finally develop into a shock at later time. More detailed analyses on formation of the shock could be found in supplementary materials.

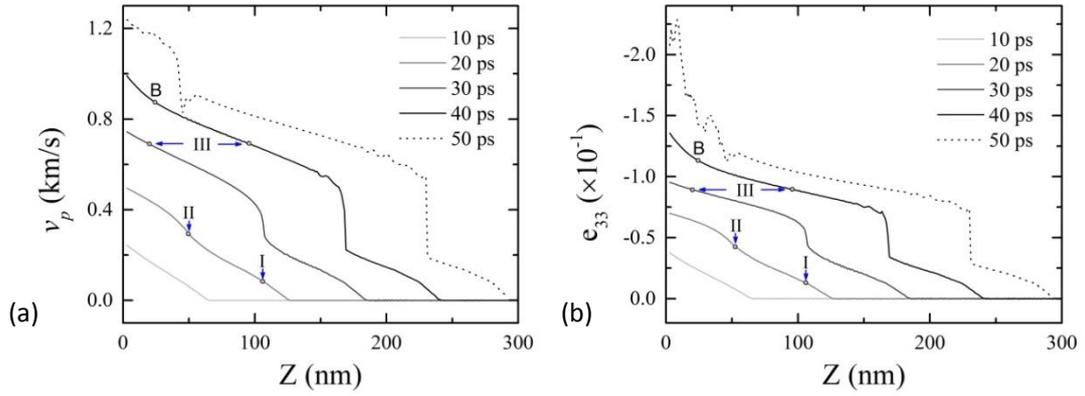

Fig. 2. Wave profile of (a) particle velocity and (b) strain for aluminum single crystal under ramp compression, with $v_{max}$ = 2.0 km/s and $t_{rising}$ = 80 ps, along [001] direction. The aluminum sample is elastically compressed until plasticity takes place at 50ps. Others are the same as Fig. 1. As shown in the figures, the knee at III is not as obvious as that at I and II. Thus, the critical strain of III, estimated from the figure (b), is not as precise as that of I and II.

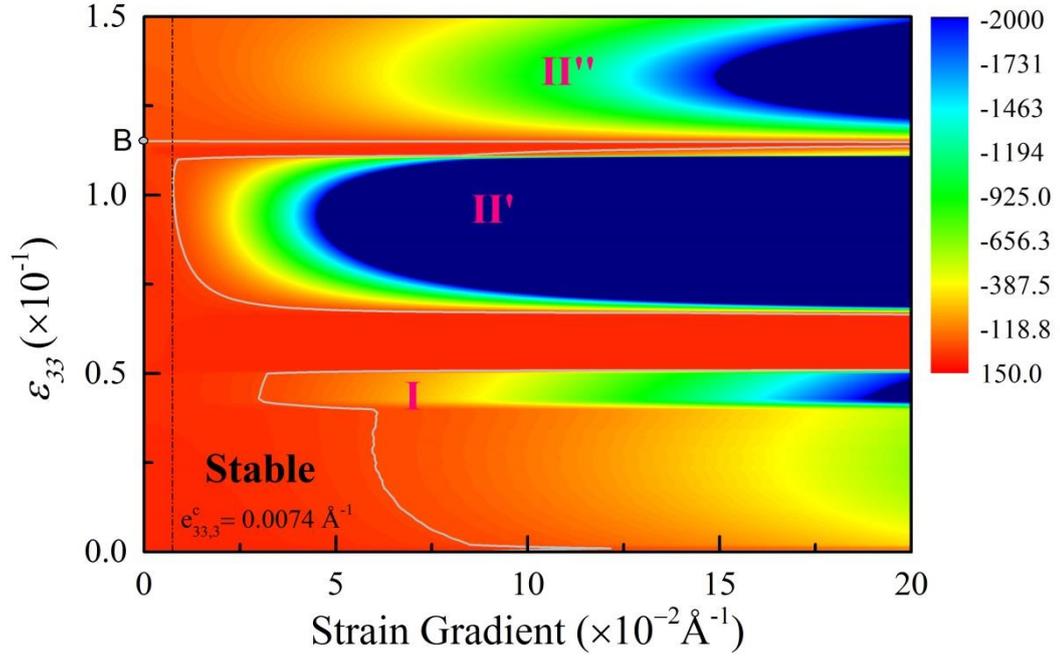

Fig. 3. $\widehat{K}_{min}$ of single crystalline Cu as a plot of uniaxial strain and strain gradient along [001] direction, where the value on the gray contour lines is zero. The critical strain for strain instability is marked by "B", which is the same as that in Fig. 1. Strain gradient instability takes place at region I and II', while strain instability occurs at region II''. The minimum strain gradient that allows the strain gradient instability to take place is marked by the black dash-dot line.

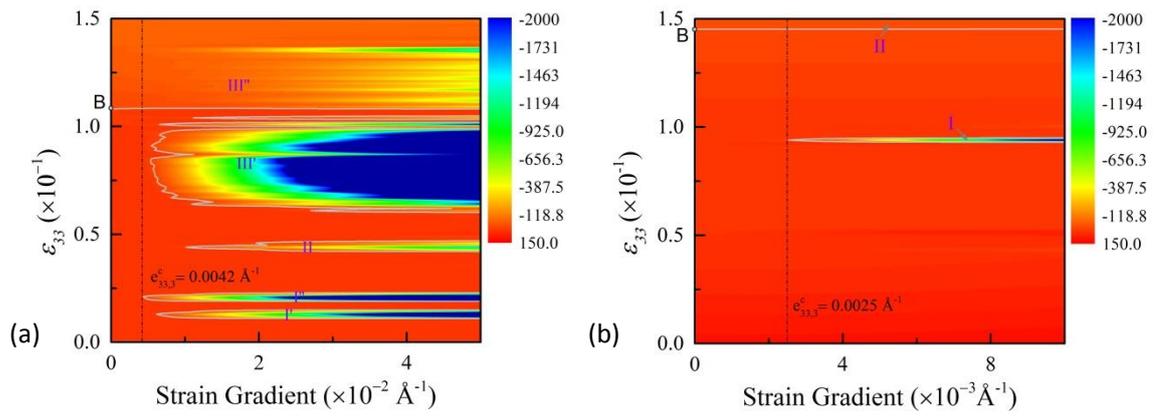

Fig. 4. $\hat{K}_{min}$ of (a) single crystalline Al and (b) Fe as a plot of uniaxial strain and strain gradient along [001] direction, where the value on the gray contour lines is zero. Others are the same as Fig. 3.